\documentclass[conference]{IEEEtran}
\usepackage{hyperref}
\usepackage{xspace}
\usepackage{todonotes}
\usepackage{subfigure}
\usepackage{algorithmic}
\usepackage{bm}
\usepackage{multirow}
\usepackage{algorithm}
\usepackage{indentfirst} 
\usepackage{amsmath} 
\newcommand{\X}{\textsc{iVantage}\xspace}
\newcommand{\Y}{RVP\xspace}
\newcommand{\Ys}{RVPs\xspace}
\usepackage{dblfloatfix} 
\ifCLASSINFOpdf
\else
\fi
\begin{document}

\title{Your Router is My Prober: Measuring IPv6 Networks via ICMP Rate Limiting Side Channels}

\author{\IEEEauthorblockN{Long  Pan\IEEEauthorrefmark{1}\IEEEauthorrefmark{2},
Jiahai Yang\IEEEauthorrefmark{1}\IEEEauthorrefmark{2}\IEEEauthorrefmark{3},
Lin He\IEEEauthorrefmark{1}\IEEEauthorrefmark{2}\IEEEauthorrefmark{3}, 
Zhiliang Wang\IEEEauthorrefmark{1}\IEEEauthorrefmark{2}\IEEEauthorrefmark{3},
Leyao Nie\IEEEauthorrefmark{1}\IEEEauthorrefmark{2},
Guanglei Song\IEEEauthorrefmark{1}\IEEEauthorrefmark{2},
Yaozhong Liu\IEEEauthorrefmark{1}\IEEEauthorrefmark{2}}
\IEEEauthorblockA{\IEEEauthorrefmark{1}
Institute for Network Sciences and Cyberspace, BNRist, Tsinghua University, Beijing, China}
\IEEEauthorblockA{\IEEEauthorrefmark{2}
Zhongguancun Laboratory, Beijing, China}
\IEEEauthorblockA{\IEEEauthorrefmark{3}
Quan Cheng Laboratory, Jinan, Shandong, China}

}

\IEEEoverridecommandlockouts
\makeatletter\def\@IEEEpubidpullup{6.5\baselineskip}\makeatother
\IEEEpubid{\parbox{\columnwidth}{
    Network and Distributed System Security (NDSS) Symposium 2023\\
    27 February - 3 March 2023, San Diego, CA, USA\\
    ISBN 1-891562-83-5\\
    https://dx.doi.org/10.14722/ndss.2023.23049\\
    www.ndss-symposium.org
}
\hspace{\columnsep}\makebox[\columnwidth]{}}

% make the title area
\maketitle

\begin{abstract}
%\boldmath
Active Internet measurements face challenges when some measurements require many remote vantage points. In this paper, we propose a novel technique for measuring remote IPv6 networks via side channels in ICMP rate limiting, a required function for IPv6 nodes to limit the rate at which ICMP error messages are generated. This technique, \X, can to some extent use 1.1M remote routers distributed in 9.5k autonomous systems and 182 countries as our ``vantage points''. We apply \X to two different, but both challenging measurement tasks: 1) measuring the deployment of inbound source address validation (ISAV) and 2) measuring reachability between arbitrary Internet nodes. We accomplish these two tasks from only one local vantage point without controlling the targets or relying on other services within the target networks. Our large-scale ISAV measurements cover $\sim$50\% of all IPv6 autonomous systems and find $\sim$79\% of them are vulnerable to spoofing, which is the most large-scale measurement study of IPv6 ISAV to date. Our method for reachability measurements achieves over 80\% precision and recall in our evaluation. Finally, we perform an Internet-wide measurement of the ICMP rate limiting implementations, present a detailed discussion on ICMP rate limiting, particularly the potential security and privacy risks in the mechanism of ICMP rate limiting, and provide possible mitigation measures. We make our code available to the community.
\end{abstract}

\section{Introduction}
Recent years have witnessed the rapid growth of IPv6 adoption.
Statistics \cite{googleipv6, akamaiipv6} show that $\sim$40\% of client systems have adopted IPv6 as of 2022.
Therefore, developing understanding of IPv6 networks by effective measurements is of great importance.
However, measuring IPv6 networks usually faces challenges. For one thing, a common challenge of Internet measurements (both IPv4 and IPv6) is the lack of vantage points. Many measurement tasks cannot be accomplished without owning one vantage point inside the target network. Therefore, a considerable cost of renting or deploying vantage points is usually necessary for an Internet-wide measurement. For another, IPv6 is considered to be more secure in many respects. For instance, the hugely expanded address space of IPv6 \cite{rfc4291, ipv6AddressDistribution} makes exhaustive scans no longer possible and poses challenges to the discovery of measurement targets such as active hosts or DNS resolvers.
The removal of the long-standing identification field (IPID) from the fixed IPv6 header \cite{rfc8200} also makes many IPID side channel-based attacks and measurements \cite{tcphijacking, idlescan, midar, augur} inapplicable to IPv6.

Similarly, unlike Internet Control Message Protocol (ICMP) for IPv4 \cite{rfc792}\protect\footnotemark, ICMP version 6 (ICMPv6) requires IPv6 nodes to limit the rate at which ICMP error messages are originated, to limit the bandwidth and forwarding costs, and reduce the risk of ICMP flooding attacks  \cite{rfc4443}.
However, while the pervasiveness of ICMP rate limiting in IPv6 networks does reduce the risk of networks being flooded with ICMP packets and makes it more difficult to perform \verb|traceroute|-based active topology discovery \cite{beholder, periphery}, it opens up new opportunities for us to measure remote IPv6 networks via ICMP rate limiting side channels.

\footnotetext{Though not required by the RFC \cite{rfc792}, many IPv4 nodes implement ICMP rate limiting \cite{asByRL, characterizeICMP, detectRL}.}

In this paper, we propose a novel ICMP rate limiting side channel-based measurement technique, \X, to measure remote IPv6 networks. With the ability to ``send'' and ``receive'' packets on remote Internet nodes by exploiting ICMP rate limiting side channels, \X can, to some extent, employ routers in remote networks as our ``vantage points'' without directly controlling them.
We apply \X to the following two tasks, both of which are difficult and even theoretically \textit{impossible} from only one local vantage point.

\textbf{Measuring Deployment of Inbound Source Address Validation}.
Source address validation reduces the risk of spoofing-based cyberattacks, such as distributed denial of service (DDoS) attacks (especially amplification attacks  \cite{ampmap,lockDoor}) and other kinds of infiltration \cite{closedDoor}, by verifying the legitimacy of the source addresses of packets. It can be divided into inbound source address validation (ISAV) and outbound source address validation (OSAV) \cite{rfc2827,closedDoor,lockDoor}.
While a majority of observed networks ($\sim$85\% prefixes and $\sim$70\% autonomous systems) have deployed OSAV according to the \textit{Spoofer} project \cite{spoofer, spooferPrj, understandingSAV}, the lack of ISAV is much more pervasive \cite{closedDoor,lockDoor}.
Therefore, it is crucial to understand the Internet-wide ISAV deployment.
However, ISAV deployment is difficult to measure because it is challenging to have vantage points in every network or autonomous system (AS).
Without any vantage points inside the target network, we cannot know whether ISAV filters our probes with spoofed source addresses.

While previous work for ISAV deployment measurements mainly relied on in-network volunteers \cite{spoofer,spooferPrj, understandingSAV} or DNS resolvers \cite{lockDoor, closedDoor}, all of our measurements can be performed from one local vantage point with better coverage, and do not rely on in-network volunteers or in-network open DNS servers.
Our measurements cover $\sim$9.7k ASes, finding that $\sim$79\% of them are (partially) vulnerable to spoofing, which is the most large-scale measurement study of IPv6 ISAV to date.

\textbf{Measuring Reachability}. As the word Internet implies, networks are interconnected to each other. We may take it for granted that every two nodes on the Internet can reach each other. However, many causes such as pervasive Internet censorship \cite{censorsurvey, globalDNSmanipulation,satellite, augur, iclab, spooky, censoredplanet, kaza, censorglobalscale, quack}, link failures \cite{pinpoint, backbonelinkfailure}, and routing failures \cite{routingpolicies,routingfailure}, can lead to loss of reachability.
Pinpointing network disruptions will help network administrators troubleshoot, and detecting censorship will help understand Internet-wide activities. As with measuring ISAV deployment, it is also challenging and even impossible to measure reachability between two remote nodes from one local vantage point without controlling any of them.
Previous work primarily used DNS \cite{globalDNSmanipulation, satellite}, virtual private networks (VPNs) \cite{iclab}, echo servers\cite{quack,censorglobalscale}, or IPID side channels \cite{spooky, augur} to perform measurements.
However, their approaches relied on many particular services with limited coverage. They focused on various kinds of Internet censorship instead of network-level reachability itself. Many methods are also inapplicable to IPv6, for instance, there is no IPID in the fixed IPv6 header, discovering targets such as echo servers is also hard because of the large IPv6 address space.

Our \X-based approach aims at measuring \textit{network-level reachability} itself. It is applicable to IPv6 networks, does not rely on global IPID counters, DNS, echo servers or VPNs, and can be done from only one local vantage point. We evaluate our method for reachability measurements with ground truth, achieving an accuracy of over 90\% and over 80\% precision and recall, respectively.

Finally, we perform an Internet-wide measurement of the implementations of ICMP rate limiting, discussing the potential risks of ICMP rate limiting and providing possible mitigation measures.

We make our source code publicly available to the community at \url{https://github.com/iVantage-NDSS23/iVantage}, which includes tools to perform Internet-wide active measurements of IPv6 ISAV and reachability measurements between two arbitrary IPv6 nodes, using methods we proposed in this paper.

\textit{Contributions.} We make the following contributions:
\begin{itemize}
  \item We propose a novel technique based on ICMP rate limiting side channels, \X, which can to some extent use 1.1M remote routers distributed in 9.5k ASes and 182 countries as our ``vantage points'' for active measurements without directly controlling them.
  \item We perform the most large-scale measurement study of IPv6 ISAV deployment to date by a novel method based on \X, from only one local vantage point, covering $\sim$50\% of all IPv6 ASes.
  \item We propose a new method based on \X for measuring the reachability (including network-level censorship) between arbitrary IPv6 Internet nodes from only one local vantage point without controlling any of them, and evaluate this method with ground truth (precision and recall of over 80\%).
  \item We perform an Internet-wide measurement of the implementations of ICMP rate limiting, provide a detailed discussion on ICMP rate limiting, reveal the potential security and privacy risks in the mechanism of ICMP rate limiting, and provide possible mitigation measures.
\end{itemize}

\section{Background and Related Work}
\label{sec:background and related work}
\subsection{ICMP Rate Limiting}
Internet Control Message Protocol (ICMP) \cite{rfc792, rfc4443} is commonly used in troubleshooting  \cite{pinpoint}, topology discovery  \cite{yarrp,beholder,flashroute,asByRL}, and network management \cite{mobilitymanagement, managementsystem}.
Well-known tools like \verb|ping|, \verb|traceroute|, and their variations are all mainly based on ICMP.
However, allowing nodes to originate ICMP messages without any restrictions can lead to waste of bandwidth and network resources, and potential risks of ICMP flooding attacks.
Therefore, in the IPv6 version of ICMP (ICMPv6) \cite{rfc4443}, IPv6 nodes are required to limit the rate at which ICMP error messages are originated. Related work \cite{asByRL,beholder} and our measurement results also demonstrate that ICMP rate limiting is more common in IPv6 networks than in IPv4.

It is worth noting that, though ICMPv6 specification recommends a token bucket-based implementation for ICMP rate limiting \cite{rfc4443}, in our measurements, it is not necessary to care about the mechanism or how the target nodes implement ICMP rate limiting. We just need to send packets, observe the rate limiting by receiving packets – and no more than that.

There has also been previous work on ICMP rate limiting.
Ravaioli et al. \cite{characterizeICMP} and Guo et al. \cite{detectRL} provided measurement studies of ICMP rate limiting. They mainly focused on the rate limiting of ICMP Echo Replies and ICMP Time Exceeded messages in IPv4 networks. However, as a mandatory function for all IPv6 nodes as specified in the RFC \cite{rfc4443}, our measurements and previous work \cite{beholder, asByRL} have demonstrated that ICMP rate limiting is more common in IPv6 networks. We also mainly focus on the rate limiting of ICMP Destination Unreachable instead of other relatively common types.

Vermeulen et al. \cite{asByRL} exploited the shared ICMP rate limiting mechanism of different router interfaces for alias resolution.
Similarly, Man et al. improved the success rate of port prediction in DNS cache poisoning attacks \cite{ccs2020} relying on insecure implementations of ICMP rate limiting for Linux-based operating systems.
Their work brings inspiration to us, revealing the insecurities in the mechanism of ICMP rate limiting.

\subsection{SAV and Its Measurement}
IP source address spoofing refers to the process of sending packets having arbitrary IP addresses as their source IP addresses.
These spoofed packets are difficult to trace and usually result in reflection and amplification attacks \cite{spoofer, ampmap, lockDoor}, flooding the victim with traffic.
For example, a spoofing-based reflection attack against Amazon Web Services in 2020 resulted in record-breaking traffic of 2.3 Tbps \cite{amsreport}.

Source address validation (SAV) was introduced and formalized in 2000 \cite{rfc2827}, serving as an essential defense against various kinds of spoofing-based attacks.
SAV mainly falls into two categories \cite{rfc2827,closedDoor,lockDoor}: 1) \textit{Outbound SAV (OSAV)} is deployed on the network egress. OSAV checks whether the source addresses of outbound traffic are indeed IP addresses within this network and discards packets whose source address does not belong to this network.
2) \textit{Inbound SAV (ISAV)} is deployed on the network ingress. Since the source address of inbound packets is unlikely to belong to the destination network, ISAV filters such packets to reduce the potential risk of spoofing-based attacks.
Measurement studies already show that lack of ISAV is much more severe than OSAV \cite{spoofer, spooferPrj,closedDoor,lockDoor}.

Many previous studies concentrated on measuring SAV deployment.
Misconfiguration-based \cite{hell, loop} approaches relied on network misconfigurations, and passive traffic-based \cite{interdomainsav, challengesav} approaches relied on inter-domain traffic. All of them have limited coverage and cannot perform Internet-wide measurements of SAV deployment.

The \textit{Spoofer} project \cite{spoofer,spooferPrj, understandingSAV} measures SAV deployment for over ten years by requiring in-network volunteers to run the \textit{Spoofer} client.
The main limitation of the \textit{Spoofer} project is that it relies on volunteers. The coverage of the \textit{Spoofer} project is also relatively small, only $\sim$1k IPv6 ASes have been measured.

Korczynski et al. \cite{lockDoor} and Deccio et al. \cite{closedDoor} measured ISAV deployment by sending spoofed and non-spoofed DNS queries and receiving queries on authoritative DNS servers. The main difference is that, Korczynski et al. scanned the whole IPv4 Internet to discover open and closed DNS resolvers, which is not feasible in IPv6 due to the large address space, while Deccio et al. utilized ``Day in the Life''  (DITL) data \cite{DITL} sponsored by the DNS Operations, Analysis and Research Center (OARC)  \cite{OARC} to find potential DNS resolvers. The main limitation of their DNS-based approaches is that for networks where there is no open DNS resolvers deployed, their approaches cannot confirm the presence of ISAV. Their coverage is also limited. The former did not measure IPv6 networks, and the latter found 3,952 IPv6 ASes lacking ISAV. Compared to their work, our \X-based approach has better coverage, and does not need one authoritative DNS server or DNS data.

Dai et al. \cite{smap} measured IPv4 ISAV via IPID and Path Maximum Transmission Unit Discovery (PMTUD) side channels. They also performed DNS-based measurements like Korczynski et al. \cite{lockDoor} and Deccio et al. \cite{closedDoor}. Their work presented a most comprehensive view of IPv4 ISAV to date. However, their IPID-based and IP fragmentation-based method are not applicable in IPv6. There is no IPID in fixed IPv6 headers. Their PMTUD-based approach requires sending probe packets with DF (don't fragment) flags, which do not exist in IPv6 fixed headers or extension headers.

\subsection{Measuring Reachability}
In most cases, any two nodes on the Internet can reach each other. However, many causes including but not limited to link failures \cite{pinpoint,backbonelinkfailure}, routing failures \cite{routingpolicies,routingfailure} and pervasive Internet censorship \cite{censorsurvey, globalDNSmanipulation,satellite, augur, iclab, spooky, censoredplanet, kaza, censorglobalscale, quack} may lead to loss of reachability. Recent research on the origin of scanning \cite{census, origin} also shows that scanning the same targets from different origins will lead to 5\% to 10\% discrepancy in the response rates. Therefore, measuring reachability between remote Internet nodes is crucial. It will help pinpoint connectivity problems, improve the reliability of Internet paths, and understand the activities in cyberspace.

Measuring reachability between two remote nodes without directly controlling one of them can be hard and even impossible. Some previous work exploited DNS \cite{globalDNSmanipulation,satellite}. However, DNS reachability cannot fully reflect the real reachability. Similarly, \textit{Quack} \cite{quack} and \textit{Hyperquack}\cite{censorglobalscale} exploited echo servers, which can only detect application-level reachability, having limited coverage. It is also hard to discover echo servers in such a large IPv6 address space. \textit{ICLAB} \cite{iclab} mainly used VPNs as vantage points. However, though \textit{ICLAB} has VPNs or in-country volunteers in 62 countries and 234 ASes, it's still not enough compared with over 190 countries and over 20k IPv6 ASes. The cost of buying and deploying VPN services is also considerable.
\textit{Spooky Scan} \cite{spooky} and \textit{Augur} \cite{augur} exploit IPID side channels to measure the connectivity disruptions. However, their methods are not applicable to IPv6 networks because there is no IPID field in the fixed IPv6 header \cite{rfc8200}.

While there exists rich literature with studies of measuring the censorship \cite{globalDNSmanipulation,satellite, augur, iclab, spooky, censoredplanet, kaza, censorglobalscale, quack}, in this paper, we will focus on the \textit{network-level reachability} itself instead of Internet censorship because Internet censorship is only one potential cause of loss of reachability. Actually, some types of censorship will also not cause loss of network-level reachability \cite{censorsurvey}.

Our \X-based approach does not rely on DNS, global IPID counter, echo servers, or VPNs and can be used to measure reachability between arbitrary IPv6 nodes from only one local vantage point.

\section{\protect\X}

\begin{figure*}[htbp]
  \centering
  \includegraphics[width=1\textwidth]{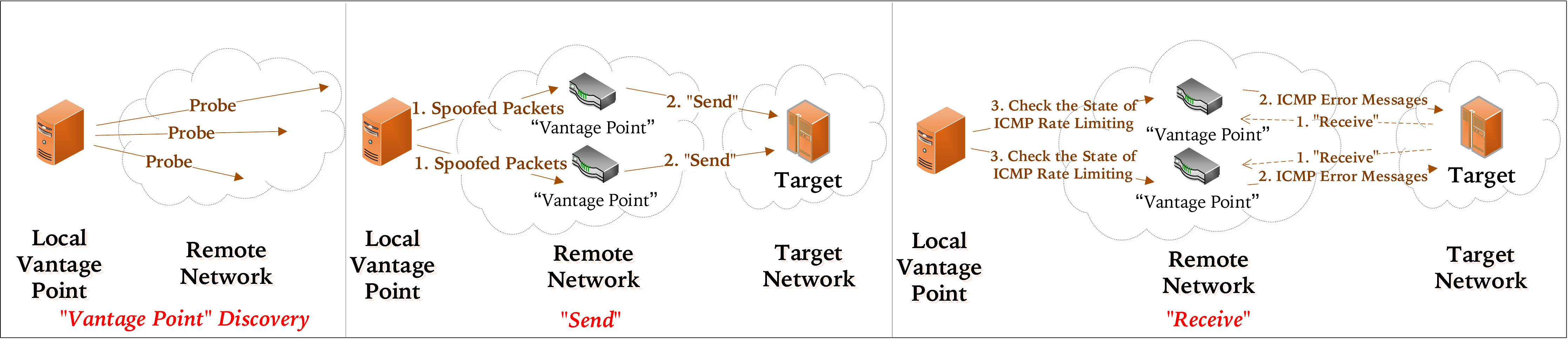}
  \caption{An Overview of \X. \X first discovers potential ``vantage points'', and then perform measurements from them with the ability to ``send'' and ``receive' on the ``vantage points''.}
  \label{fig:technique}
  \vspace{-0.3cm}
\end{figure*}

    A common challenge with active network measurements is the lack of vantage points.
    Without the ability to \textit{send} or \textit{receive} packets on a vantage point within a remote network, many measurement tasks cannot be completed.
    However, the reality is that it is not possible to have vantage points in all networks.
    Therefore, we propose an ICMP rate limiting-based technique, \X, to solve this challenge, which consists of the following steps: finding remote \textit{``vantage points''} first and then having them \textit{``send''} and \textit{`` receive''} packets.

    \textbf{``Vantage Point'' Discovery.}
    \X's first step is to find targets that may serve as our ``vantage points'', even though they are not vantage points in the true sense. They are mainly routers in remote networks, which are not under our control. We refer to them as remote ``vantage points'' (\Ys).
    All the IPv6 nodes that will originate ICMP messages in response to particular probes can be used as \Ys. In \S \ref{sec:rvp discovery}, we describe a simple but effective method for discovering \Ys.

    \textbf{``Send''.}
    After discovering the available \Ys, we need to control the \Ys to send probes for active measurements, which is the most crucial part. However, since we have no control over the \Ys, we can only induce them to send probes. An intuitive way to do this is to use spoofed source addresses to send packets to the \Ys that require the \Ys to send responses, e.g., ICMP Echo Requests, TCP SYN. Upon receiving the probes, the \Ys will send replies to the spoofed source addresses, just as if we were asking the \Ys to send probes as we intended. Inducing \Ys to send replies is not necessarily helpful because, in general, the replies sent by the \Ys do not have any observable effect on the measurement targets. However, if the \Ys send replies that cause the measurement targets to send ICMP error messages, these replies will have an observable effect in the form of \textit{different states of ICMP rate limiting}. Distinguishing between these different states can help us to measure targets.

    \textbf{``Receive''.}
    In most measurements, there is also a need to capture packets from an \Y. However, without access to the \Y, capturing its received packets is difficult. Fortunately, for some measurements, it is generally not necessary to capture all packets on the \Y. We only care about whether (or even when) it receives a particular type of packets. The different states of ICMP rate limiting can help confirm whether (or even when) the packets are received. If the \Y receives a considerable number of packets that would cause it to send ICMP error messages, it can be observed that the \Y's ICMP rate limiting is triggered. Observing this difference makes it possible to know if the \Y receives these packets, just like capturing packets on the \Y.

    Figure \ref{fig:technique} provides an overview of \X. Suppose we want to perform measurements from a vantage point inside a remote network, but we do not have such a vantage point.
    \X will first find \Ys in the remote network.
    Then, the key is how to let \Ys send or capture packets as we wish without directly controlling them.
    As mentioned earlier, if we want the \Ys to send packets to the target, we send packets to the \Ys using the target's source address. Then the \Ys will send replies to the target, just as if we asked \Ys to send probes.
    Similarly, it is difficult to know whether \Ys can receive the packets sent by the target.
    However, suppose the packets that the target sends to \Ys will induce \Ys to originate ICMP error messages (e.g., these packets are sent to unreachable IP addresses). In that case, we can know whether \Ys have received these replies by checking the state of ICMP rate limiting on the \Ys.
    
    \X mainly aims at measuring filtering policy and network connectivity of remote networks with the ability to ``\textit{send}'' and ``\textit{receive}'' packets on the \Ys. Admittedly, \Ys can not perform all kinds of measurements as real vantage points. In the rest of this paper, we apply \X to two different but both challenging measurement tasks.
    In the measurements of ISAV deployment (\S \ref{sec:measuring isav deployment}), we focus primarily on how to ``\textit{receive}'' packets on \Ys ;
    in the measurements of reachability (\S \ref{sec:measuring reachability}), we rely on \Ys to ``\textit{send}'' and ``\textit{receive}'' both.
    Both of these two tasks theoretically require in-network vantage points.
    See how \X uses others' routers as our ``vantage points''
    and make the impossible possible.

    \section{``Vantage Point'' Discovery}
    \label{sec:rvp discovery}

    To fully exploit ICMP rate limiting side channels, we need to collect a set of IP address pairs (hereafter referred to as \textit{data pairs}).
    The data pairs consist of IP address pairs in the form of $\langle$\textit{target, periphery}$\rangle$: by sending an ICMP Echo Request (i.e., \verb|ping|) to the \textit{target}, we can receive an ICMP error message from the \textit{periphery}, which is usually different from the \textit{target}.
    Since such ICMP error messages (mostly ICMP Destination Unreachable in our measurements) are usually originated by the last hop router on the path to the \textit{target}, we call it \textit{periphery}.
    We borrow the definition of \textit{periphery} from \cite{periphery}, which refers to the last hop router connecting end hosts. These peripheries are actually the remote ``vantage points'' (\Ys) that will be used in the subsequent measurements.
    
    \begin{figure}[htbp]
      \centering
      \includegraphics[width=0.48\textwidth]{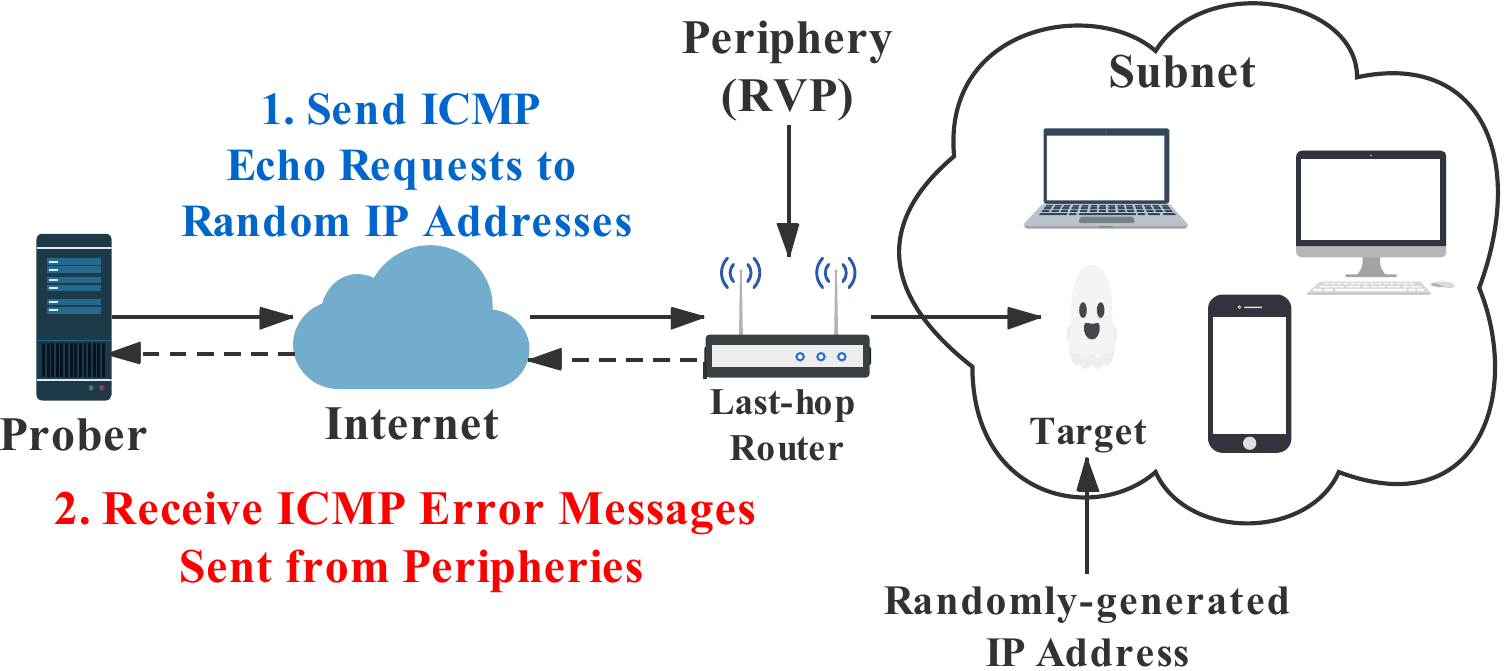}
      \caption{Process of ``Vantage Point'' Discovery. The local prober sends lots of ICMP Echo Requests to the randomly-generated addresses and capture ICMP error messages.}
      \label{fig:periphery}
    \end{figure}
    
    \begin{figure*}[h!t]
      \centering
      \includegraphics[width=1.0\textwidth]{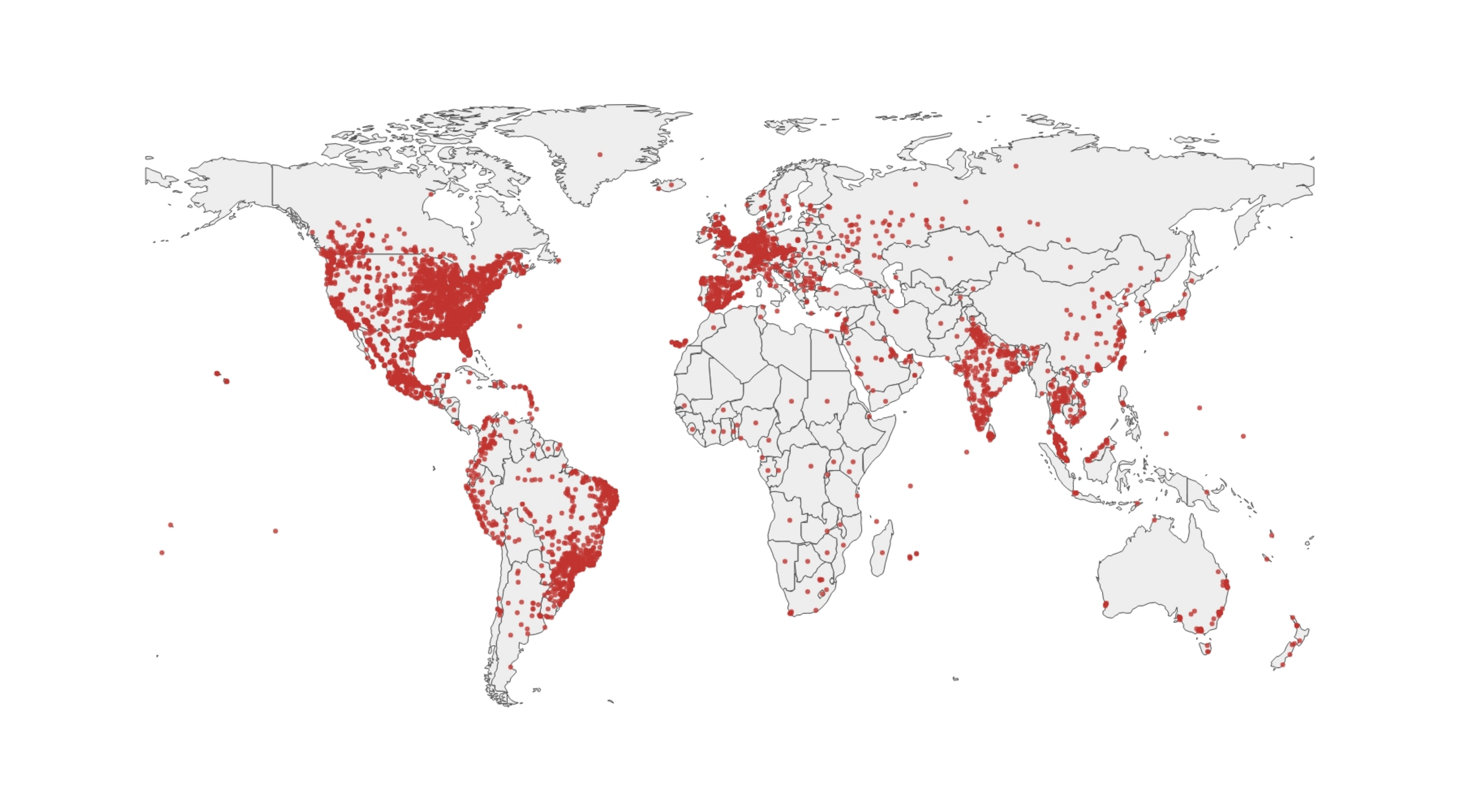}
      \caption{Geographical Distribution of Our ``Vantage Points''\protect\footnotemark}
      \label{fig:vp}
    \end{figure*}

    \footnotetext{Some locations may be inaccurate.}

    To collect the data pairs, we implement a stateless scanner like many existing high-speed probers \cite{ZMap, yarrp, flashroute}. 
    In consideration of the large address space of IPv6 networks, we cannot perform an exhaustive scan. Therefore, unlike other scanners that usually send probes to the IP addresses given, our scanner performs scans according to the $\sim$100k global IPv6
    BGP prefixes present in the BGP routing table (maintained by RouteViews \cite{RouteViews}), and generates targets by the following method \cite{prefixrotation, xiangli}:
    For every BGP prefix, we combine different /64 prefixes and random interface identifiers (the last 64 bits of IPv6 addresses).
    For instance, for BGP prefix 2000:1234::/40, we may have 2000:1234:00\textbf{00:0000}\textcolor{gray}{:adef:4983:19d2:3f12} first (the gray segments are generated randomly),
    and then 2000:1234:00\textbf{00:0001}\textcolor{gray}{:fed1:4f12:0894:349d} is the next.
    The last target
    may be 2000:1234:00\textbf{ff:ffff}\textcolor{gray}{:1f3e:175a:4159:4de1}. In practice, we randomize the probing sequence.
    When the scanner keeps probing each prefix, we capture \textit{ICMP error messages} in response to our probes at the same time. The ICMPv6 specification \cite{rfc4443} only requires IPv6 nodes to limit the rate of originating ICMP error messages\protect\footnotemark, and our measurements (see \S \ref{sec:discussion}) also reveal that the rate limiting of other types of ICMP packets (like ICMP Echo Reply) is less obvious and thereby more difficult to be exploited as side channels.
    Since the targets are randomly generated, they are very likely
    to trigger a great many ICMP error messages.
    We can easily extract both \textit{target} and \textit{periphery} from the ICMP error messages we received because it is required for ICMP error messages to quote the invoking packet (including, of course, its destination address) \cite{rfc4443}.
    
    \footnotetext{Though not required by the RFCs \cite{rfc4443, rfc792}, it's common for Internet nodes to limit the rate of originating ICMP Echo Replies \cite{asByRL, characterizeICMP, detectRL}. We also exploit the rate limiting of ICMP Echo Replies to perform supplemental ISAV measurements.}

    Figure \ref{fig:periphery} illustrates how the process works on one topology model. There is one periphery (usually CPE router or base station) before one customer subnet. Our probes are sent to the subnets, most of them are destined to unreachable IP addresses, triggering ICMP error messages sent from the peripheries.

    We limit the sending rate and use the multiplicative group of integers modulo $n$ \cite{gauss,ZMap} to randomize the probing sequence, refraining from sending probes to one network in succession.
    In addition, we do \textit{not} send that many packets for each BGP prefix; instead, we stop scanning as soon as we find enough (e.g., 50) data pairs in a BGP prefix to limit the number of packets we send. We will also stop scanning if we are still unable to find any data pairs after sending a sufficient number (e.g., 1M) of packets to that network.

    We repeated the scan several times in three months from a local vantage point on the university campus. Finally, we found \textbf{1,118,817} distinct data pairs that will be used as \Ys in subsequent measurements.
    Among all the ICMP error messages we received, 87.2\% are ICMP Destination Unreachable, and the rest are ICMP Time Exceeded. It's very surprising to receive many ICMP Time Exceeded messages because for all the packets we sent, we set the hop limit (a.k.a. Time to Live, TTL) to 64. It may be attributed to the routing loops or other kinds of misconfigurations \cite{xiangli}.

    Figure \ref{fig:vp} shows the distribution of remote ``vantage points'' (using \verb|GeoLite2| \cite{geolite2} to locate IP addresses). Our ``vantage points'' are distributed in \textbf{182} different countries, \textbf{29,679} BGP prefixes and \textbf{9,498} ASes.

    \begin{figure*}[htb]
      \begin{center}
        \includegraphics[width=0.95\textwidth]{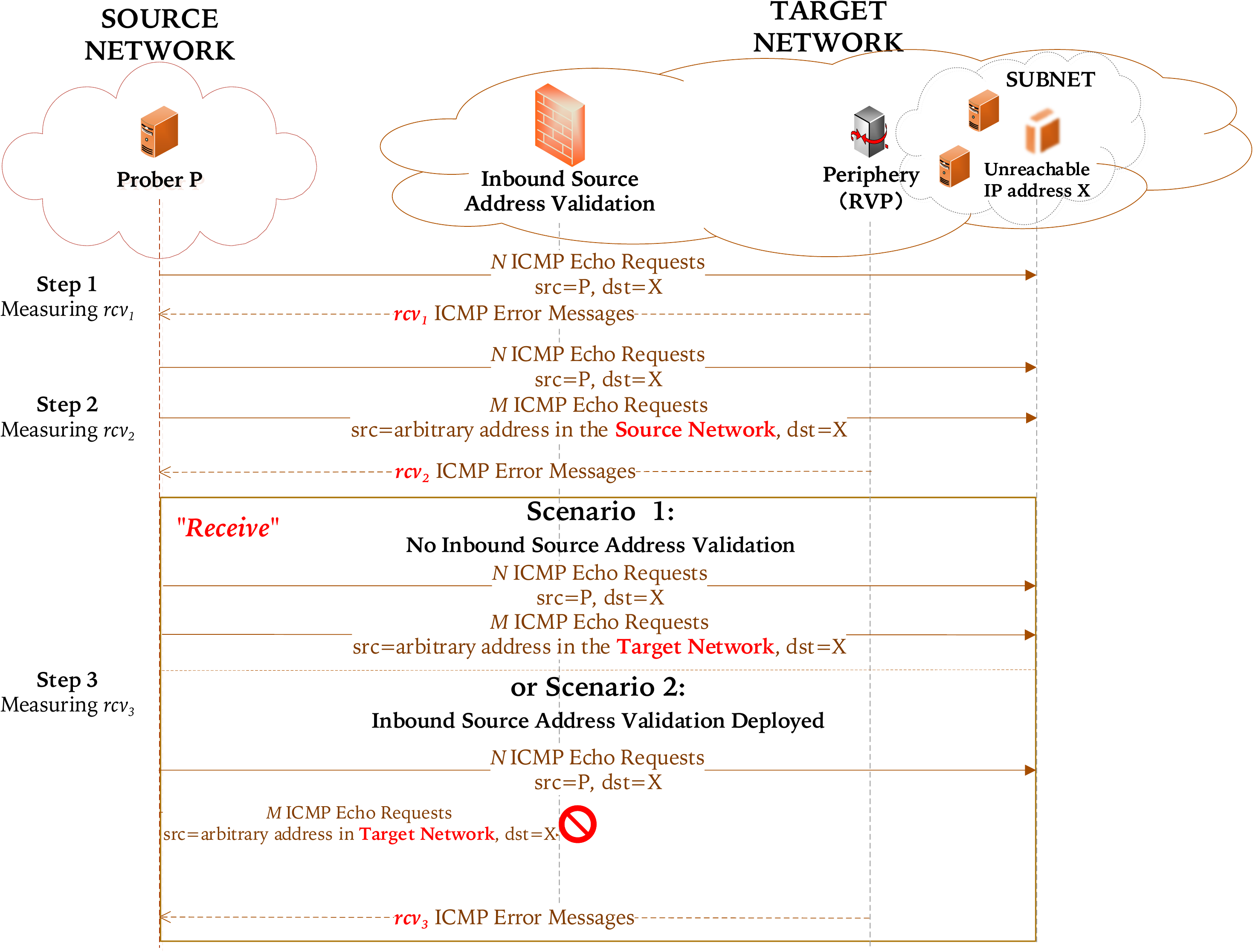}
        \caption{Methodology of Measuring Inbound Source Address Validation Deployment}
        \label{fig:sav}
      \end{center}
    \end{figure*}

    \section{Measuring ISAV Deployment}
    \label{sec:measuring isav deployment}
    In this section, we apply \X to the measurement of ISAV deployments. The goal is to identify whether the target network (or BGP prefix, AS) has deployed ISAV to filter spoofed inbound packets.
    The main challenge is how to know whether the spoofed packets are filtered by ISAV, i.e., whether the \Ys can \textit{``receive''} the spoofed packets, which, fortunately, \X can easily do, as mentioned earlier.

    Each target network will fall into one of the following three categories:
    \begin{itemize}
      \item \textbf{ISAV Deployed}: ISAV is deployed in the target network, which usually happens on the border of the target network.
      \item \textbf{Vulnerable to Spoofing}: ISAV is not deployed on the path toward the \Y in the target network. The target network is vulnerable to IP address spoofing attacks.
      \item  \textbf{Uncertain}: The deployment of ISAV in the target network cannot be determined.
    \end{itemize}

    \subsection{Methodology}
  
    Figure \ref{fig:sav} illustrates the method for measuring ISAV deployment via ICMP rate limiting side channels.
    We have a local prober P in the source network. In contrast, we have a pair of IP addresses in the target network, the target (i.e., X in Figure \ref{fig:sav}) and the periphery (i.e., the \Y), which are collected previously.
    We focus on measuring three values $rcv_1$, $rcv_2$, and $rcv_3$:
    
    \begin{itemize}
      \item $rcv_1$: The number of ICMP error messages (sent from \Y) P receives after sending $N$ ICMP Echo Requests (i.e., \verb|ping|) to the \textit{target}. We refer to these packets as \textit{probe packets}.
    
      \item $rcv_2$: Similar to $rcv_1$, but besides sending the same probe packets, P also sends $M$ ICMP Echo Requests to the \textit{target} simultaneously with a spoofed source address that belongs to the network of P\protect\footnotemark. We call such packets \textit{noise packets}.
    
            \footnotetext{It is not absolutely necessary to spoof the source address here. We spoof in order not to receive ICMP error messages caused by noise packets.}
    
      \item $rcv_3$: Similar to $rcv_2$, but the spoofed source address of the noise packets belongs to the target network.
    \end{itemize}

    Our method is based on measuring and comparing the averages of the above three values.

    $\bm{rcv_1}$ vs. $\bm{rcv_2}$.
    Theoretically, as long as the \Y implements (global) ICMP rate limiting, we have $rcv_1 > rcv_2$.
    This is because noise packets exacerbate \Y's ICMP rate limiting.
    If there is $rcv_1\approx rcv_2$, we can know that the \Y doesn't implement ICMP rate limiting or only implements very loose (or non-global) ICMP rate limiting, and therefore cannot be a good \Y in ISAV measurements. In practice, we will ignore these \Ys.

    $\bm{rcv_2}$ vs. $\bm{rcv_3}$.
    The process of measuring $rcv_2$ and $rcv_3$ is very similar. The only difference is the network to which the spoofed source address of noise packets belongs. If it belongs to the network of P, the spoofed source address is hard to be identified by the target network; if it belongs to the target network, it will be detected and filtered by ISAV (if deployed).
    Hence, if there is $rcv_2 < rcv_3$, we can infer that the target network has filtered the noise packet, and therefore, ISAV is deployed.
    In this case, we usually have $rcv_1\approx rcv_3$.
    This is because there is no difference between the processes of measuring $rcv_1$ and $rcv_3$ if all the spoofed-source noise packets are filtered.
    Similarly, if there is $rcv_2\approx rcv_3$, it makes no difference which network the spoofed source address of the noise packets belongs to, suggesting no ISAV on the path toward the target network. Actually, our measurements find that if ISAV is not deployed, we often find $rcv_2 > rcv_3$ instead of $rcv_2\approx rcv_3$.  This may be because remote Internet nodes prefer to reply to ICMP Echo Requests sent from their proximate Internet nodes \cite{origin}, so noise packets with spoofed source addresses within the target network trigger a more obvious ICMP rate limiting.

    $\bm{rcv_1}$ vs. $\bm{rcv_3}$.
    As mentioned before, if there is $rcv_1 > rcv_3$, it means that noise packets with spoofed source addresses within the target network are not filtered, so ISAV does not exist; if $rcv_1\approx rcv_3$, ISAV exists because the noise packets are filtered, so there is no essential difference between $rcv_1$ and $rcv_3$.

    \subsection{Inferring ISAV Deployment}
    We infer the ISAV deployment by comparing the averages of these three values: $rcv_1$, $rcv_2$, and $rcv_3$. As mentioned above:
    \begin{itemize}
      \item For networks with ISAV: $\overline{rcv_1}\approx \overline{rcv_3} > \overline{rcv_2}$.
      \item For networks without ISAV: $\overline{rcv_1} > \overline{rcv_2} \approx \overline{rcv_3}$ or $\overline{rcv_1} > \overline{rcv_2} > \overline{rcv_3}$.
      \item For networks without ICMP rate limiting (or too loose, too strict, non-global ICMP rate limiting):  $\overline{rcv_1} \approx \overline{rcv_2} \approx \overline{rcv_3}$.
    \end{itemize}
    
    In practice, we infer the ISAV deployment based on simple but very effective rules. We introduce a factor $0 < \lambda < 1$ to avoid potential interference from fast-changing network environments, and we can be sure of $a < b$ only if $a < \lambda \times b$.
    \begin{itemize}
      \item If $\overline{rcv_3} < \lambda\times \overline{rcv_1}$, then ISAV is not deployed;
      \item Else if $\overline{rcv_2} < \lambda\times \overline{rcv_3}$, then ISAV is deployed;
            % \item Else if $$rcv_3$ < \lambda\times\verb|rcv1|$, then ISAV doesn't exist;
      \item Else, the deployment of ISAV cannot be determined.
    \end{itemize}
    
    It's possible to have both $ \overline{rcv_3} < \lambda \times \overline{rcv_1}$ and $\overline{rcv_2} < \lambda \times \overline{rcv_3}$.
    However, our measurements show that it rarely occurs (less than 5\%).
    For these special cases, we compare the values of $\overline{rcv_3} / \overline{rcv_1}$ and $\overline{rcv_2} / \overline{rcv_3}$, and then rely on the smaller one.
    All three values will be measured multiple times to avoid the interference from uncertain network states and possible packet loss. We only compare the averages of these three values.

    We simply compare the averages of $rcv$'s instead of using other more statistically rigorous methods because we find the $rcv$'s (i.e., $rcv_1$, $rcv_2$, $rcv_3$) we measured on one same \Y in multiple measurements are relatively consistent in our early experiments. On average, 88.3\% of them are same as their modes, i.e., the $rcv$ value that appears most often in multiple measurements.

    \begin{figure}[htbp]
      \centering
      \includegraphics[width=0.48\textwidth]{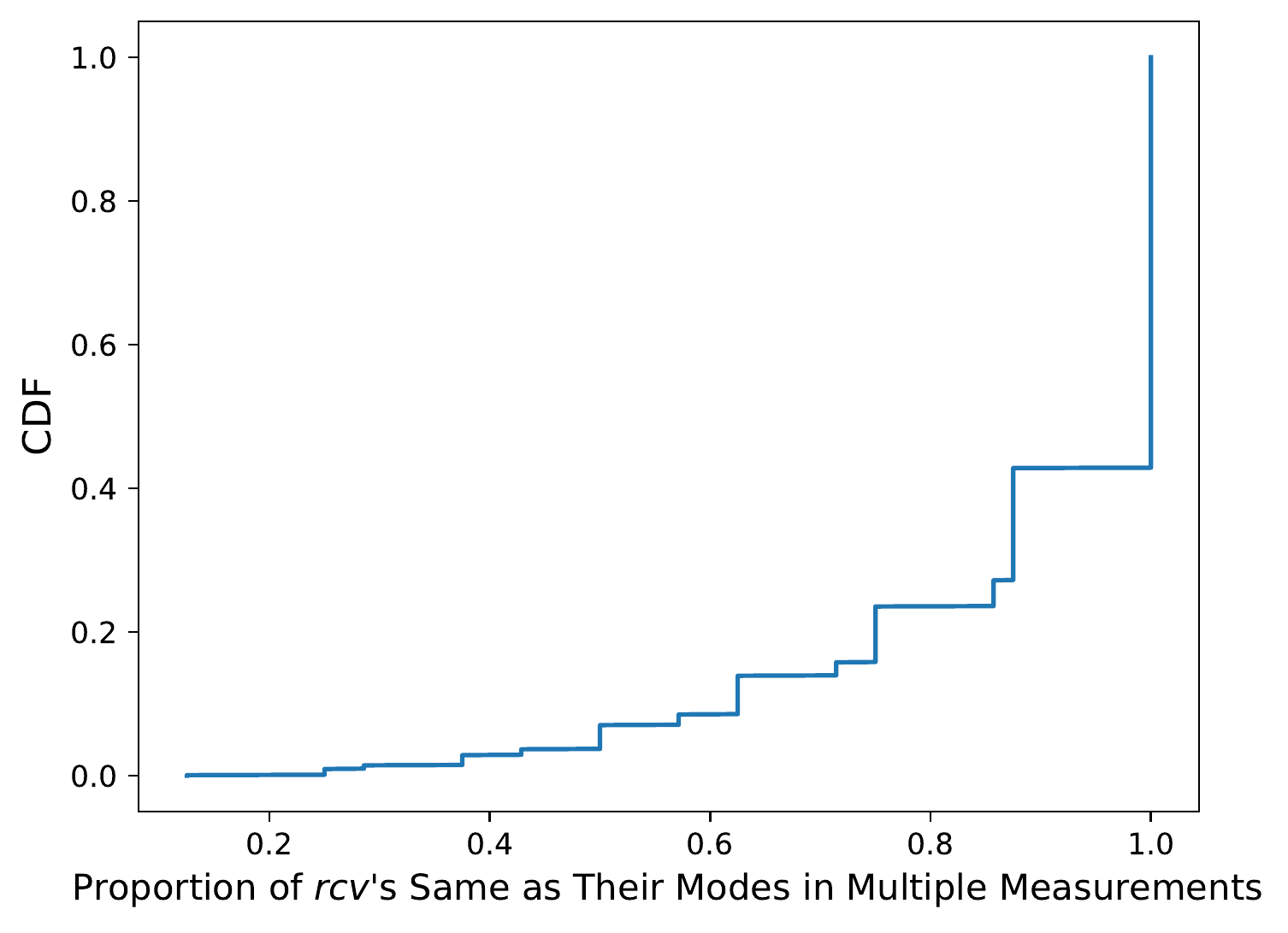}
      \caption{Distribution of Proportions of $rcv$'s that are Same as Their Modes in Multiple Measurements. Results are from Early ISAV Measurements on $\sim$2k \Ys.}
      \label{fig:mode}
    \end{figure}

    \subsection{A Large-scale Measurement}
    \subsubsection{Experiment Setup}
    We aim at measuring the deployment of ISAV in all the advertised BGP prefixes.
    We choose BGP prefix-level granularity because some (large) ASes may contain BGP prefixes with different ISAV policies.
    Note that our method can be used to measure networks of any size. We measure from a single vantage point on the campus network. Our measurements last for about 3 months.
    
    \textbf{Target Selection.}
    After \Y discovery, we collected data pairs distributed in $\sim$30k BGP prefixes.
    For every BGP prefix, we will choose an appropriate \Y within it.
    It is common to have many \Ys within one target BGP prefix,
    so we prefer the \Ys implementing ICMP rate limiting that are neither too strict nor too loose.
    If the ICMP rate limiting of an \Y is too strict, e.g., replying with only one ICMP error message no matter how many ICMP Echo Requests it receives, it is difficult to observe the different states of its ICMP rate limiting.
    Similarly, suppose the ICMP rate limiting is too loose, e.g., replying with 50 ICMP error messages in response to 50 ICMP Echo Requests that we send out quickly. In that case, it is also difficult for us to observe the difference.
    Sending more ICMP Echo Requests, e.g., 100, may help us observe more obvious ICMP rate limiting of \Ys, but it may have a negative impact on the target network.
    Therefore, \Ys that implement moderate ICMP rate limiting are prioritized.
    In practice, we will ignore the \Ys with $rcv_1=N$ or $rcv_1=1$ because the ICMP rate limiting is too strict or loose unless such \Ys are the only ones in the target network.
    
    \textbf{How Many Packets to Send?}
    To determine the values of $N$ and $M$, we first consult some documentations of well-known router manufacturers like Cisco, Juniper, Huawei and HPC \cite{juniper, cisco, huawei, hpc}. We found that both the implementations and default thresholds of ICMP rate limiting vary considerably with different devices, OS versions and manufacturers. Some of them implement a token bucket-based (e.g., a maximum of 10 tokens and new token placed at an interval of 100 milliseconds \cite{hpc}) as recommended by the RFC \cite{rfc4443}, while there are also many devices that rate limit the packets per second (pps) of ICMP packets (ranging from 1 pps \cite{juniper} to 100 pps \cite{huawei}). Based on the findings, we believe it's most reasonable to choose a value lower than the maximum pps we found (i.e., 100 pps) as the value of $N$ and choose a bigger value as $M$ to make $N+M$ apparently higher than 100.

    Actually, there is a trade-off when we choose the number of packets to send. More packets trigger more obvious ICMP rate limiting but may bring ethical issues or lead to waste of resource. Sending few packets is more friendly to target networks but may result in less observable rate limiting. We randomly choose $\sim$1000 \Ys to perform early experiments.

\begin{figure}[htb]
      \begin{center}
        \includegraphics[width=0.48\textwidth]{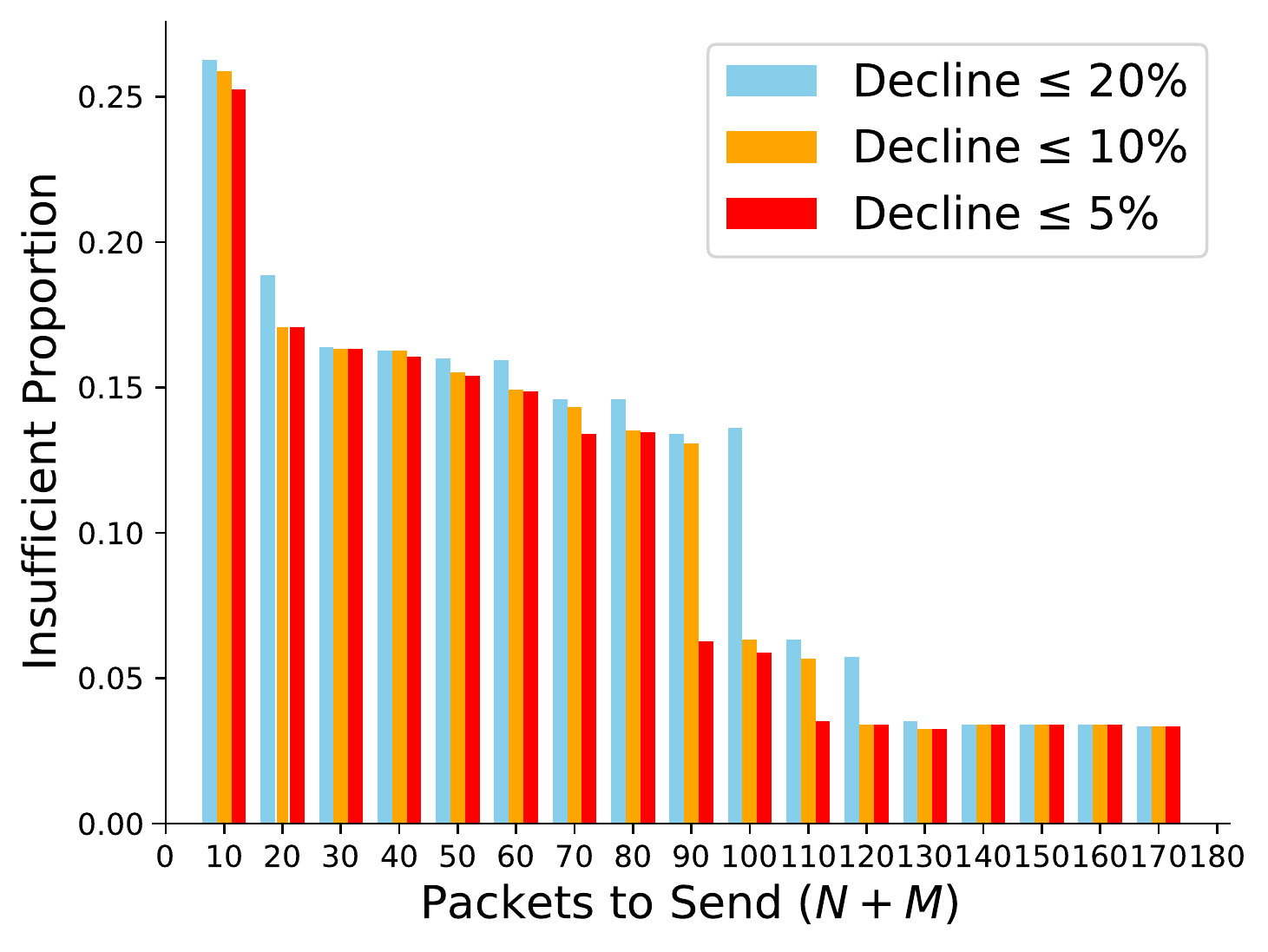}
        \caption{Insufficient Proportion of \Ys When Sending Different Numbers of Packets}
        \label{fig:ngpro}
      \end{center}
    \end{figure}

    First, we test how many packets (i.e., $N+M$) are sufficient enough to trigger observable rate limiting. If $N+M$ packets still fail to trigger obvious rate limiting, our method doesn't work. Therefore, we require $N+M$ packets to result in a relatively obvious rate limiting (e.g., 5\%, 10\% or 20\% decline in the replies we received); otherwise, the packets are \textit{insufficient}. Figure \ref{fig:ngpro} shows the proportion of insufficient cases when sending different numbers of packets. Packets with a total number of 30-100 remain insufficient for about 15\% \Ys, where we can only observe less than 5\%, 10\% or 20\% decline. 
    The proportion of insufficient cases decreases significantly when the packet number exceed 100 and 120. 130 packets are sufficient in more than 95\% cases. The experimental results are consistent with our expectations according to preliminary findings. Therefore, a total of $\sim$130 packets seem to be adequate for our measurements.

    Secondly, we test how many probe packets and noise packets can make the impact of noise packets more observable. Since we have already total number of $\sim$130 packets (i.e., $M+N$) are already sufficient in most cases, what we need to do is to determine the ratio of probe packets ($N$) to noise packets ($M$). Taking into account possible packet loss in more complicated network environments when it comes to Internet-wide measurements, we define $M+N=150$ and then try different values of $M/N$.
    We define the \textit{observability} as the decline in the proportion of the replies we can receive after we send noise packets (i.e., $1 - rcv_{after} / rcv_{before}$) and then measure the observability of different ratios of $M$ to $N$ (i.e., $M/N$). Table \ref{tab:obs} implies that bigger $M/N$ will result in better observability and the observability stabilizes at $\sim$0.4 when $M/N \ge 2$. We finally choose $M/N=2$ instead of 2.5 or larger because we think too many noise packets and too few probe packets cannot help us characterize the behavior of rate limiting well. 
    
    \begin{table}[]
      \caption{Observability of Different Ratios of Noise Packets to Probe Packets}
      \begin{center}
        \begin{tabular}{cccccc}
          \hline
          \multicolumn{1}{l|}{}    & \multicolumn{5}{c}{$M/N$}                  \\ \cline{2-6} 
          \multicolumn{1}{c|}{}    & 0.5    & 1      & 1.5    & 2      & 2.5    \\ \hline
          \multicolumn{1}{c|}{$M$ (Noise Packets)} & 50     & 75     & 90     & 100    & 107     \\
          \multicolumn{1}{c|}{$N$ (Probe Packets)} & 100    & 75     & 60     & 50     & 43    \\ \hline
          \textbf{Observability}   & 0.3171 & 0.3650 & 0.4072 & 0.4192 & 0.4272 \\ \hline
        \vspace{-1.0cm}  
        \end{tabular}
        \label{tab:obs}
      \end{center}
      \end{table}

    We finally choose $N=50$ and $M=100$ in our measurements, which is in line with our previous expectations with $N$ being a value smaller than 100 and $N+M$ being bigger than 100. The research community can also choose other values to balance those trade-offs when performing similar measurements.

    \textbf{Choice of $\bm{\lambda}$}.
    The choice of $\lambda$ can be very hard because of limited ground truth of ISAV deployment. The existing measurement results of ISAV are mostly private because of potential risks of those no-ISAV networks being attacked. Actually, the main problem is how obvious is obvious enough? For instance, $\overline{rcv}=0$ obviously proves the presence of rate limting, but how about $\overline{rcv}=0.2N, 0.4N,...0.7N$?
    After comparing different $\lambda$ in small-size ground truth by DNS-based measurements like \cite{lockDoor,closedDoor} and considering the evaluation result of the reachability measurement with ground truth (see Table \ref{tab:con_res} in \S \ref{sec:measuring reachability}), we finally choose $\lambda = 0.6$ in the measurements. That is, when inferring the ISAV deployment by the rules introduced earlier, we confirm that $a < b$ only if $a < 0.6 \times b$.
    Experiments show that the value of $\lambda$ in the range of 0.5 to 0.7 will not obviously affect the results. After all, in a general sense, a $\sim$40\% decline in the \textit{average} number of ICMP messages is believed to be self-evident enough to reveal the triggering of ICMP rate limiting.

    \begin{figure*}[htbp]
      \centering
      \subfigure[Error Message-based Measurements]{\includegraphics[width=0.45\textwidth]{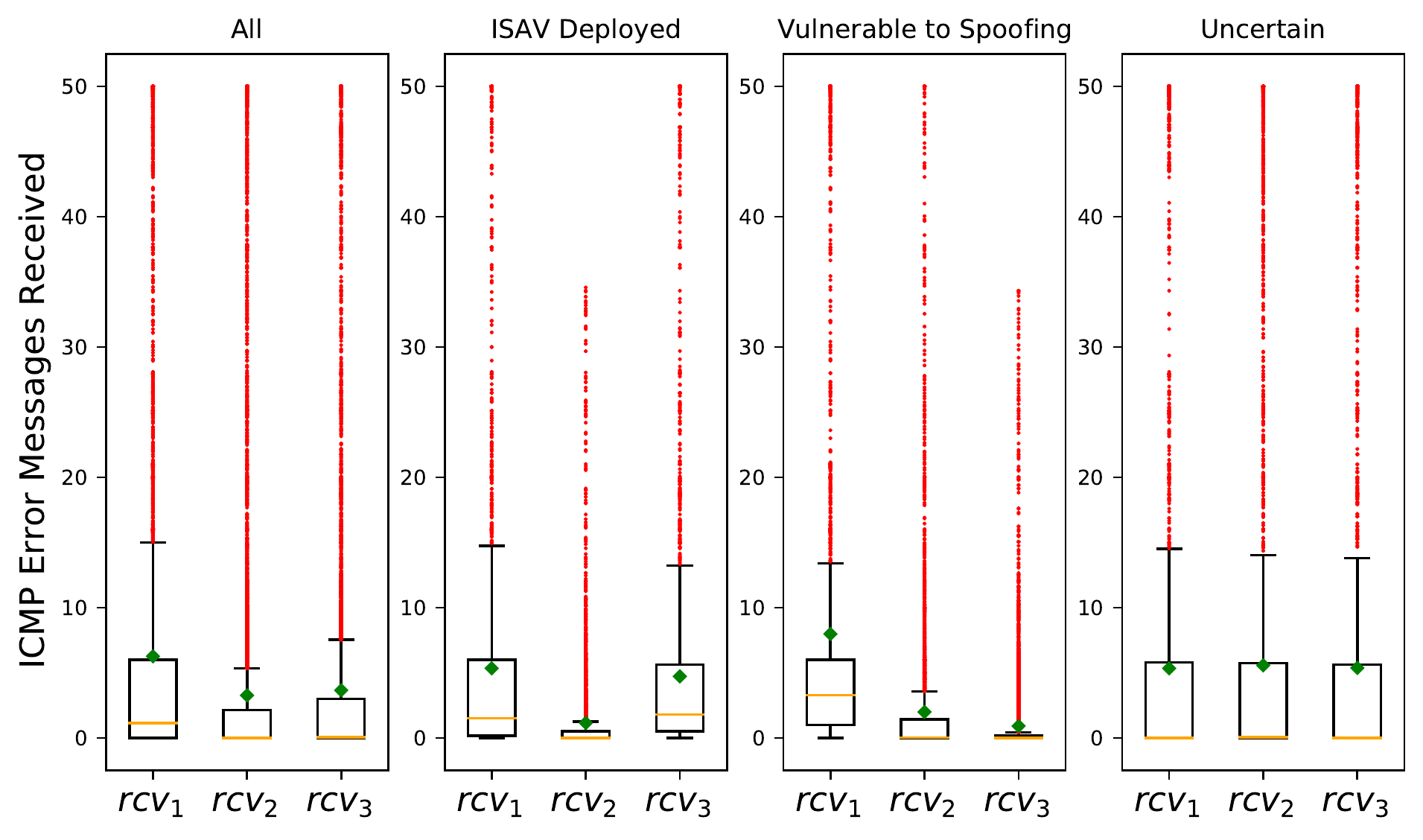}}
      \subfigure[Echo Reply-based Supplemental Measurements]{\includegraphics[width=0.45\textwidth]{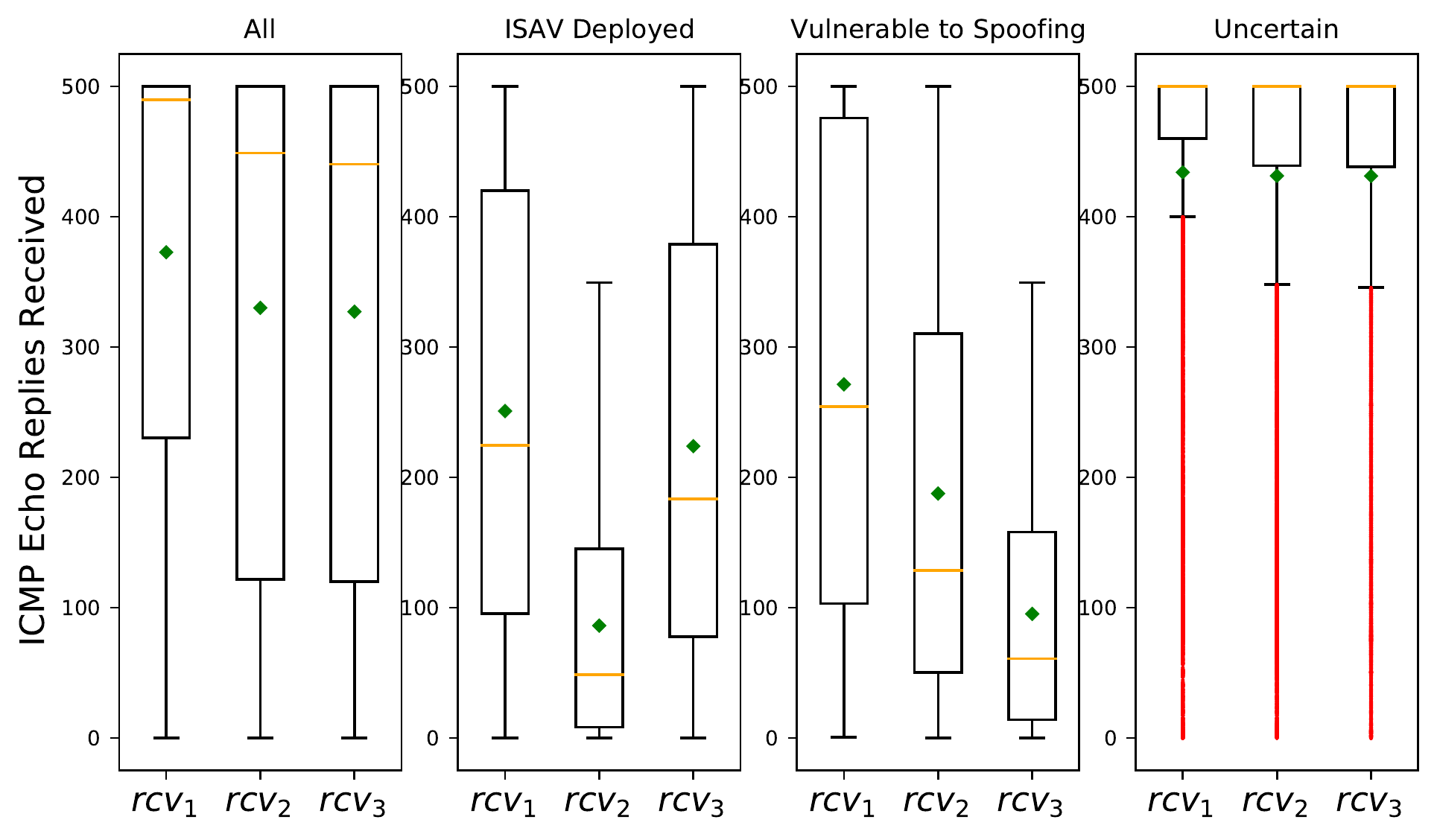}}
      \caption{Distributions of $rcv_1$, {$rcv_2$} and {$rcv_3$} under Different Categories. (Orange Lines: Medians. Green Squares: Averages)}
      \label{fig:avg_rcv}
    \end{figure*}
    
    \begin{table*}[htbp]
      \begin{center}
        \caption{Measurement Results of ISAV Deployment}
        \begin{tabular}{cccc}
          \hline
          \multicolumn{2}{c}{\textbf{BGP Prefixes}} & \multicolumn{2}{c}{\textbf{Autonomous Systems}}                                                    \\ \hline
          \textbf{Vulnerable to Spoofing}           & 20830 (75.66\%)                                 & \textbf{Vulnerable to Spoofing} & 6520 (67.37\%) \\
          \textbf{ISAV Deployed}                    & 6702 (24.34\%)                                  & \textbf{ISAV Deployed}          & 2020 (20.87\%) \\
          \textbf{Uncertain}                        & 19319                                           & \textbf{Inconsistent}           & 1137 (11.75\%) \\
          \textbf{Total}                            & 46851                                           & \textbf{Total}                  & 9677           \\ \hline
        \end{tabular}
        \label{tab:sav_res}
        \vspace{-0.4cm}
      \end{center}
    \end{table*}

    \begin{figure}[htbp]
        \centering
        \includegraphics[width=0.5\textwidth]{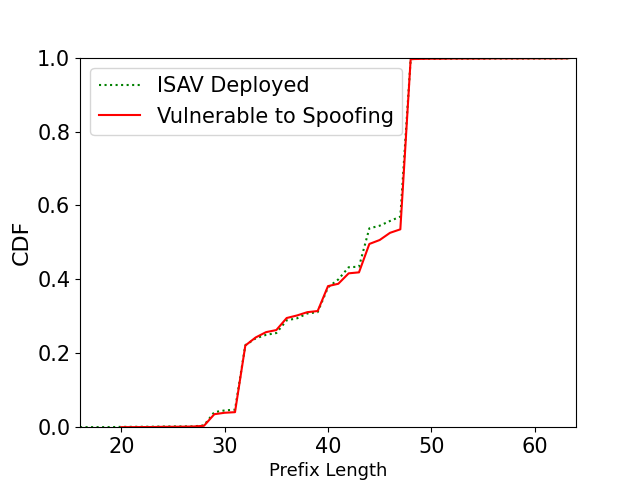}
        \caption{Prefix Lengths vs. ISAV Deployment}
        \label{fig:cdf_len}
    \end{figure}
    
    \begin{figure}[htbp]
        \centering
        \includegraphics[width=0.5\textwidth]{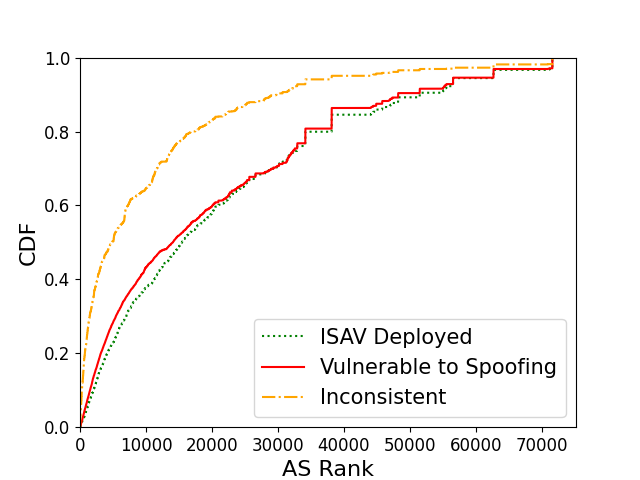}
        \caption{AS Ranks vs. ISAV Deployment}
        \label{fig:cdf_asrank}
    \end{figure}

    \begin{figure*}[htbp]
      \centering
      \includegraphics[width=1.0\textwidth]{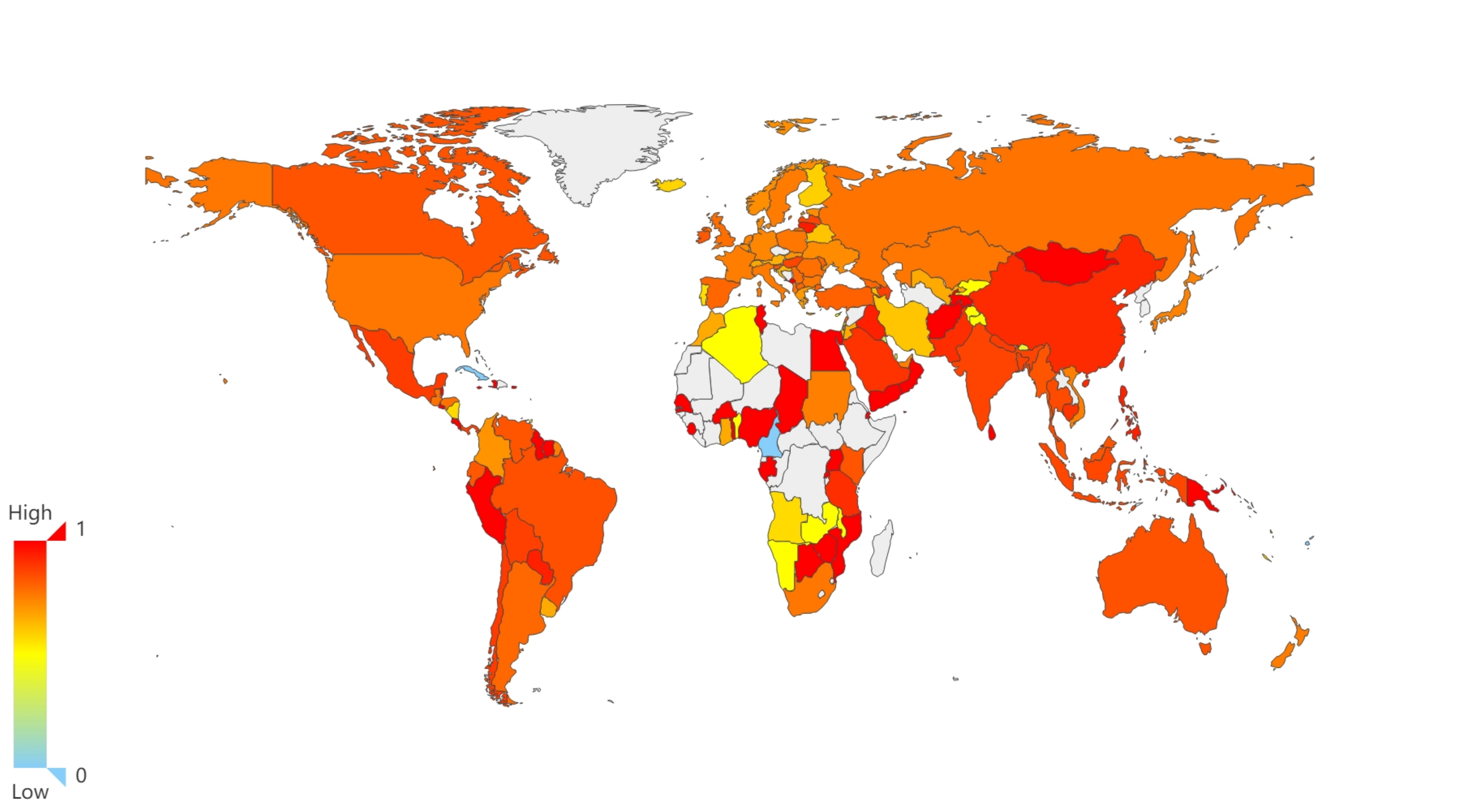}
      \caption{Fraction of Vulnerable to Spoofing vs. All ASes Per Country}
      \label{fig:sav_map}
    \end{figure*}

    \textbf{Spoofed Source Addresses.}
    As for the spoofed source address, we choose an arbitrary address with the same /80 prefix as the local vantage point as the spoofed source address in the source network. Actually, we can choose any address in the source network, as long as the spoofed source packets can be sent to the Internet. Similarly, we choose an arbitrary IP address with the same /124 prefix as the \Y as the spoofed source address in the target network, which ensures that the spoofed source address is indeed in the same target network as the \Y.

    \textbf{Measurements.}
    We continuously measure the values of $rcv_1$, $rcv_2$, and $rcv_3$ of each BGP prefix. However, we do \textit{not} measure these three values for a prefix consecutively.  If we measure $rcv_1$, $rcv_2$, and $rcv_3$ for a prefix without interruption, the ICMP rate limiting for an \Y may be triggered repeatedly in a short time.
    Instead, we will first measure $rcv_1$ for the first prefix and then $rcv_1$ for the second prefix. After measuring all the values of $rcv_1$, we will measure $rcv_2$ for the first prefix. For every prefix, we will measure $\sim$10 times and calculate the averages of $rcv_1$, $rcv_2$, and $rcv_3$ to avoid potential random errors or statistical bias.

    \textbf{Supplemental Measurements.}
    For those BGP prefixes that remain uncertain after measurement, we also perform additional measurements using the rate limiting of ICMP Echo Replies, though, the rate limiting of ICMP Echo Replies is much looser and challenging to observe (see \S \ref{sec:discussion}). That is to say, we directly send ICMP Echo Requests to \Ys instead of unreachable IP addresses and capture the ICMP Echo Replies instead of ICMP error messages.
    To enrich the targets, we add active IP addresses within those uncertain BGP prefixes from the IPv6 hitlist (6.3M addresses) collected by Gasser et al. \cite{hitlist}. However, we prioritize the \Ys we previously discovered because many of the IP addresses from the hitlist are core routers instead of the peripheries. They implement extremely loose ICMP rate limiting, which is difficult for us to observe the difference. We let $N=M=500$ in the ICMP Echo Reply-based supplemental measurements. Table \ref{tab:rl_type} shows that $N=500$ is sufficient to trigger observable rate limiting of ICMP Echo Replies for $\sim$65\% of the targets. Larger $N$ and $M$ may help us observe more obvious rate limiting, but it may lead to ethical issues since numbers larger than 500 are too large.

    \subsubsection{Results}
    Figure \ref{fig:avg_rcv} shows the distribution of $rcv_1$, $rcv_2$, and $rcv_3$ as measured in the ICMP error message-based measurements and in the supplemental ICMP Echo Reply-based measurements.
    For all target networks, we find that $rcv_1$ is larger than $rcv_2$ and $rcv_3$ because the noise packets make a difference by triggering or exacerbating the ICMP rate limiting of the \Ys.
    We infer the ISAV deployment and then draw the box plots of the three values under three different categories.
    We find that the results are as expected. For networks where we infer that ISAV is deployed, we have $rcv_1 \approx rcv_3 > rcv_2$, while for networks lacking ISAV, $rcv_1 > rcv_2 \approx rcv_3$ or $rcv_1 > rcv_2 > rcv_3$, which also shows that our simple rules do work.
    In addition, there are networks for which we cannot infer their ISAV deployment because $rcv_1 \approx rcv_2 \approx rcv_3$.
    
    Table \ref{tab:sav_res} is the summary of the ISAV measurement results. In comparison with relatively well-deployed OSAV \cite{spoofer, spooferPrj},
    the lack of ISAV is alarming.
    Our measurements, the most large-scale study of IPv6 ISAV to date, cover 27,532 BGP prefixes, 9,677 ASes, and 186 countries, and identify $\sim$2$\times$ more IPv6 ASes lacking ISAV than the state-of-the-art measurements \cite{closedDoor}.
    
    The results show a pervasive absence of ISAV. Intuitively, one may think that some large BGP prefixes, high-rank (according to AS Rank \cite{asrank}) ASes or developed countries are more likely to deploy ISAV, but the results are counterintuitive. Figure \ref{fig:cdf_len} and Figure \ref{fig:cdf_asrank} show no clear correlation between the ISAV deployment and the sizes (ranks) of the BGP prefixes (ASes). However, it is noteworthy that Figure \ref{fig:cdf_asrank} shows that high-rank ASes tend to have different ISAV policies, which is as we expected. It is not reasonable for some very large (or high-rank) AS to have only one ISAV policy. Figure \ref{fig:sav_map} provides a global map of ISAV deployment.

    \subsubsection{Validation}
    As there is no ground truth of Internet-wide ISAV deployment, we directly or indirectly validate the measurement results from the following aspects:
    
    \textbf{Validation by DNS Resolvers.}
    We replicate the previous work \cite{lockDoor,closedDoor} on measuring ISAV based on DNS resolvers by registering a dedicated domain name and deploying an authoritative DNS server.
    We scanned the entire IPv4 address space to discover all open DNS resolvers by adding an extensive probe module to \verb|ZMap| \cite{ZMap} and found 1,950,278 potential open DNS resolvers, of which 1,212,406 could forward queries to our authoritative DNS server.
    For each potential IPv4 open DNS resolver, we obtained its IPv6 address using DNS based on the method introduced by Hendriks et al. \cite{ipv6ddos} and finally found 4,692 potential IPv6 addresses of open DNS resolvers.
    We repeat the measurements in \cite{lockDoor,closedDoor}, and we find that for BGP prefixes identified by our measurements as vulnerable to spoofing, 87.94\% of them were confirmed by ISAV measurements based on DNS resolvers.
    The conflicting portion of the results could be attributed to different ISAV policies in one BGP prefix. 
    The open DNS resolvers we find are mainly in surprisingly large BGP prefixes, with 4,692 IPv6 addresses belonging to BGP prefixes with an average length of 23.91.
    Thus, different ISAV policies may be deployed in different subnets of such a large network prefix.

    \begin{figure}[hbp]
      \centering
      \includegraphics[width=0.48\textwidth]{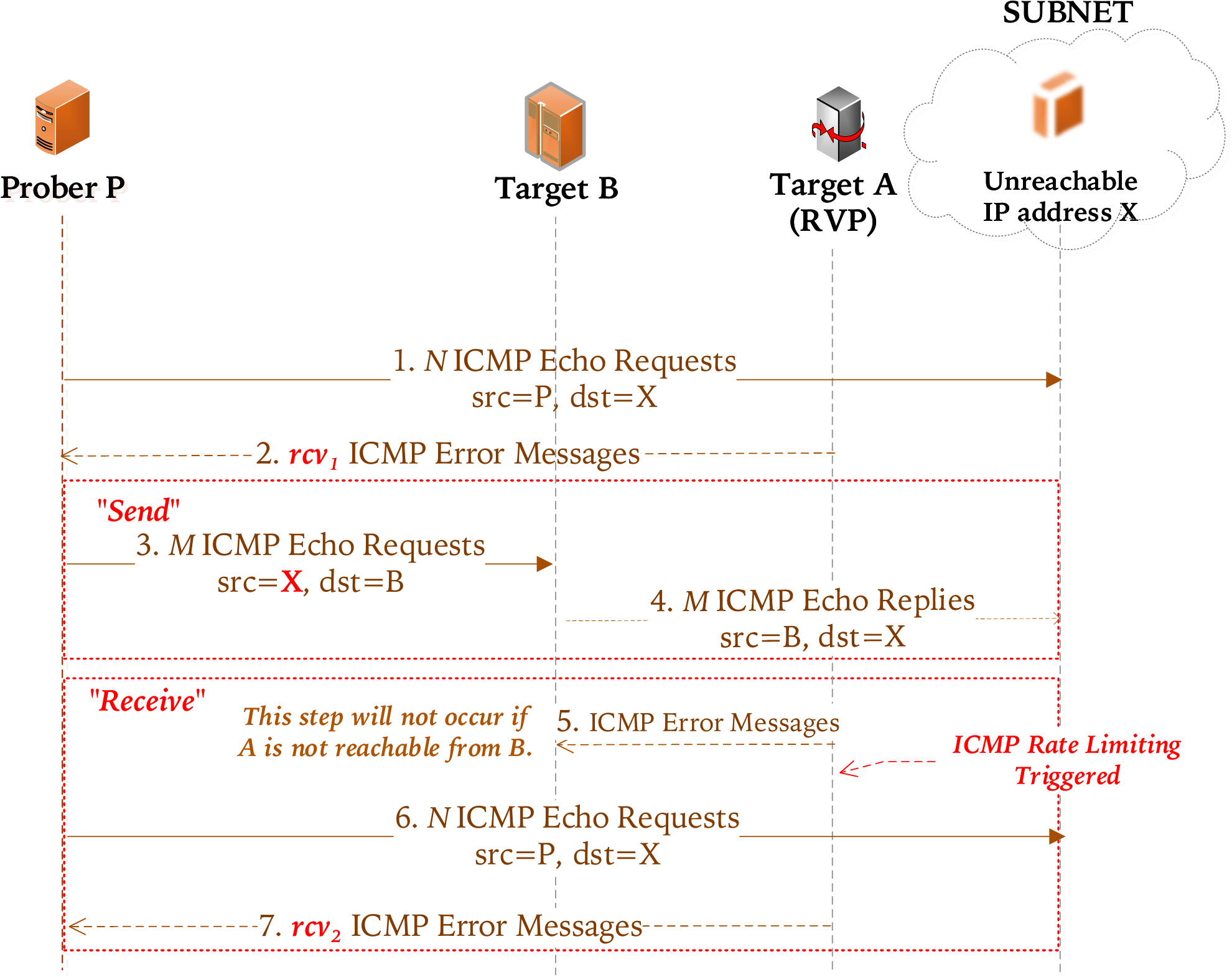}
      \caption{Methodology for Measuring Reachability between Internet Node A and B. We ``send'' probes using Target B, and then ``receive'' packets on Target A. Though only one \Y is shown in the figure, actually, Target B is also used as a ``vantage point'' to ``send'' our probes.}
      \label{fig:reachability}
    \end{figure}
    
    \textbf{Comparison with Previous Work.}
    Luckie et al. \cite{spoofer, spooferPrj} found that 68.6\%-74.2\% of tested IPv6 ASes lacking or partially lacking ISAV deployment.
    Their results are compelling because \textit{Spoofer} relies on in-network volunteers.
    Our measurements also found that 67\% -79\% of tested IPv6 ASes lacking or partially lacking ISAV deployment.
    Deccio et al. \cite{closedDoor} found that $\sim$50\% of tested IPv6 ASes lacking ISAV based on DNS resolvers.
    Their number is relatively small because there are no open DNS resolvers deployed in some ASes.
    Similar to our results, Dai et al. \cite{smap} found 69.8\% ASes lacking ISAV.

    \subsection{Limitations}
    The main limitation of our approach is that it will not work well in networks where we cannot find one \Y or the \Ys (or the active IP addresses from the hitlist \cite{hitlist}) we find do not implement (global) ICMP rate limiting, or just implement very loose or strict ICMP rate limiting. However, all of these obstacles will not result in misjudgments because those networks will remain uncertain. 
    
    \section{Measuring Reachability}
    \label{sec:measuring reachability}
    In this section, we apply \X to the measurements of reachability between two remote nodes on the Internet, neither of which is under our control.
    Measuring the reachability between two remote nodes requires us to \textit{``send''} and \textit{``receive''} packets without controlling them, which can be done by \X as mentioned before.

    \subsection{Methodology}
    
    To measure the reachability between two Internet nodes, namely A and B, we first need to find an \Y. It will be easier if either A or B is already an \Y that we have discovered in the process of ``vantage point'' discovery.
    However, suppose that neither A nor B is an \Y. In that case, we refer to the previously collected \Ys to find an appropriate \Y as close to A or B as possible (i.e., a \textit{proxy RVP}). Usually, we can find at least one appropriate proxy \Y for A or B since the 1.1M \Ys are widely distributed across $\sim$30k BGP prefixes, $\sim$9.5k ASes, and 182 countries. A proxy \Y can be a compelling proxy for A (or B) because they are usually in the same BGP prefix (or even a smaller prefix). 
    The assumption here is that the loss of reachability between two very close Internet nodes is unlikely to occur. Assuming that A is an \Y, our method consists of the following steps, as shown in Figure \ref{fig:reachability}.

    \begin{enumerate}
      \item Local prober P sends $N$ ICMP Echo Requests (very rapidly) to an unreachable IP address X (i.e., the \textit{target} in the data pairs collected previously), which will result in ICMP error messages sent from A.
    
      \item P receives $rcv_1$ ICMP error messages from A. In most cases, $rcv_1 < N$ because of the ICMP rate limiting.
    
      \item P sends $M$ ICMP Echo Requests to B with a spoofed source address, which is the same unreachable IP address we use in step 1.
    
      \item B replies with $M$ ICMP Echo Replies sent to that unreachable IP address. Note that ICMP Echo Replies are much more difficult to trigger the ICMP rate limiting, so B replies with $M$ ICMP Echo Replies.
    
      \item Since these replies are sent to an unreachable IP address, the last hop router A replies with several ICMP error messages. \textit{If A is unreachable from B, this step does not happen.}
    
      \item At approximately the same time, P again sends $N$ ICMP Echo Requests to that unreachable IP address without spoofing the source address.
    
      \item P receives $rcv_2$ ICMP error messages.
    \end{enumerate}

    \subsection{Inferring Reachability}
    We infer reachability by comparing $rcv_1$ and $rcv_2$. If B can reach A, A will receive ICMP Echo Replies sent from B to an unreachable IP address, which will trigger ICMP rate limiting on A. So in step 7, P cannot receive as many ICMP error messages as before.
    If B cannot reach A, A will not receive the messages sent by B. Then step 5 will not happen. Therefore, its rate limiting will not be triggered, and we can receive as many ICMP error messages as before.
    We also infer the reachability based on the average of repeated experiments to avoid possible packet loss or other interference in a single experiment.

    \subsection{Measurements \& Evaluation}
    \subsubsection{Ground Truth}
    Almost any two nodes on the Internet can reach each other, so it is difficult to find some unreachable pairs of IP addresses in such a large IPv6 address space.
    We first obtain 10M active IPv6 addresses by the method proposed by Song et al. \cite{DET}.
    Then, we deploy two vantage points A and B located in two different continents, approximately 8,000 km from each other.
    By scanning those 10M addresses using \verb|ZMapv6| \cite{ZMapv6} from both vantage points, we finally find 149 consistently abnormal IP addresses as ground truth (``needles in a haystack!''). These addresses are reachable from A, but packets sent from them cannot reach B. This could be attributed to link failures or inter-domain routing failures because most of these addresses are in relatively small ASes (with an average AS rank \cite{asrank}
    of 12059.7) and are also geographically distant (geolocated by \verb|GeoLite2| \cite{geolite2}) from the vantage point B (only 6/149 are in the same continent as the vantage point B).
    In addition, we select additional 851 normal IP addresses reachable from both A and B as a control group, adding up to 1,000 IP addresses. After all, abnormal IP addresses with reachability problems are much fewer than those normal IP addresses if we perform real measurement on the Internet.

    \subsubsection{Measurements}
    We perform measurements from vantage point A, aiming at distinguishing these 149 abnormal IP addresses that are unconnected with vantage point B from other 851 normal IP addresses. From the data pairs we previously collected, we find 3 \Ys in the same /56 prefix of B as the proxy \Ys.
    To prevent continuous ICMP rate limiting on the same router, these 3 \Ys are used in rotation, and also with an interval of five minutes.
    
    With the method introduced earlier,
    we continuously record how many ICMP error messages the proxy \Y can receive in step 7 (i.e., $rcv_2$) because it is not necessary to repeat step 1 and 2 in a multi-target measurement. We send $N=50$ probe packets and $M=100$ noise packets. We measure every $rcv_2$ 6 times and then calculate the average.

    \begin{figure}[htb]
      \centering
      \includegraphics[width=0.45\textwidth]{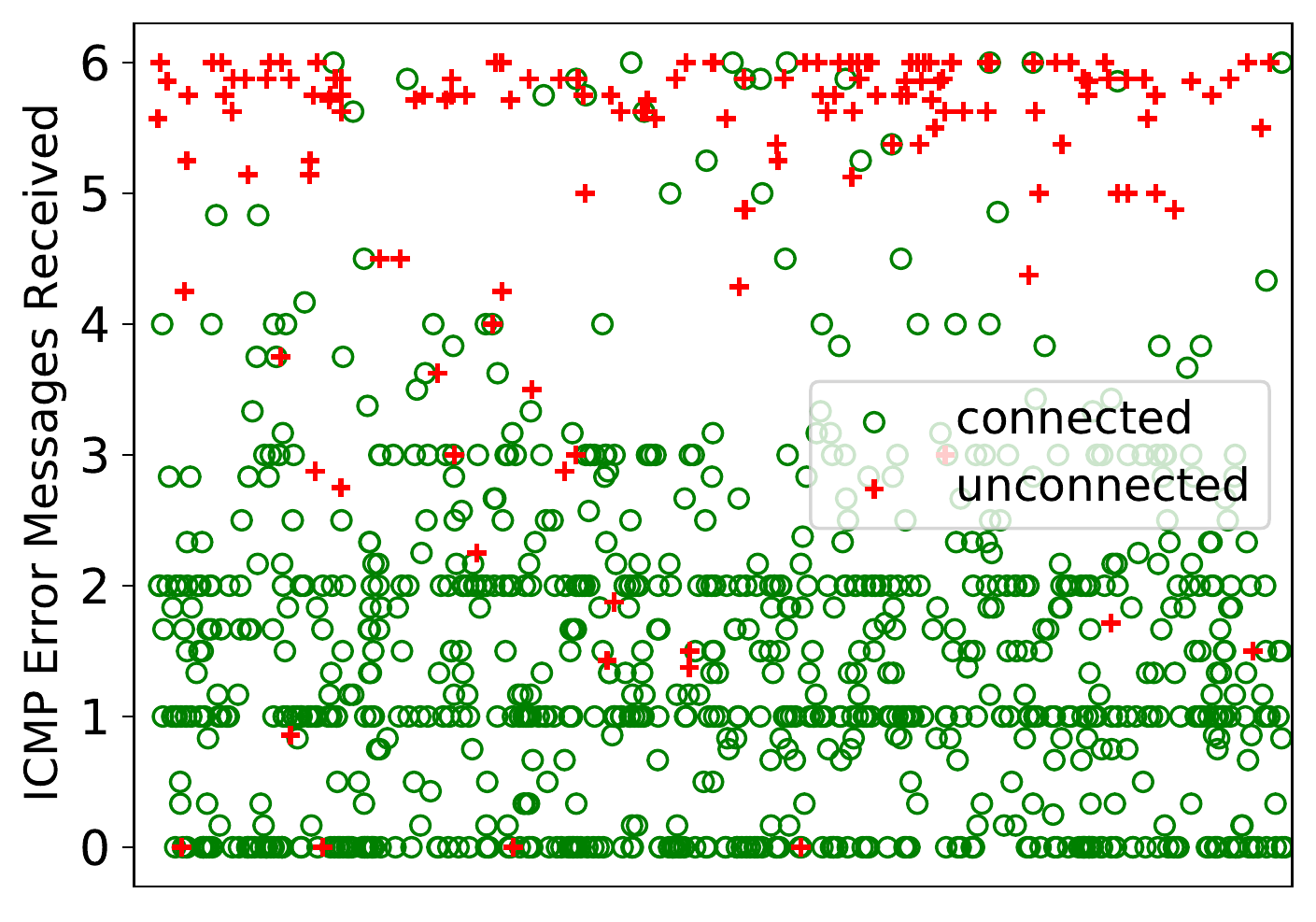}
      \caption{Average Numbers of ICMP Error Messages Received ($\overline{rcv_2}$) for Connected and Unconnected IP Addresses}
      \vspace{-0.2cm}
      \label{fig:con_val}
    \end{figure}
    
    \begin{figure}[htb]
      \centering
      \includegraphics[width=0.47\textwidth]{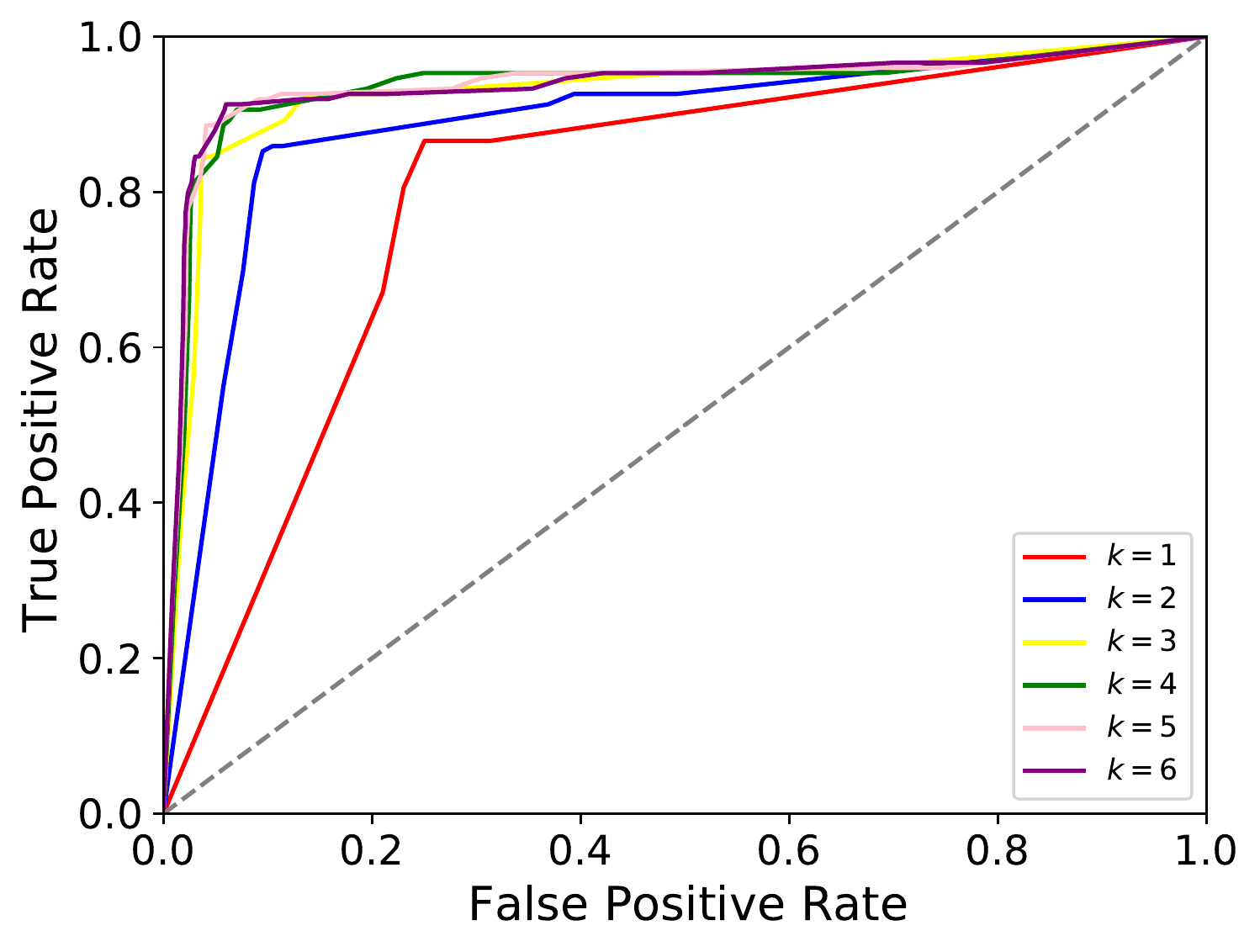}
      \caption{ROC Curve of Our Method for Measuring Reachability\protect\footnotemark}
      \label{fig:roc}
    \end{figure}

    \subsubsection{Evaluation}
    Figure \ref{fig:con_val} shows the average values of $rcv_2$ when measuring the
    reachability between vantage point B and connected or unconnected IP addresses.
    Note that for unconnected IP addresses, we obviously receive more ICMP error messages (specifically 5.131 vs. 1.472 on average) as expected.
    
    \begin{table}[ht]
      \centering
      \caption{Experimental Results of Our Reachability Measurements ($k = 6$)}
      \begin{tabular}{ccccc}
        \hline
        {$\bm{\lambda}$} & {\textbf{Precision}} & \textbf{Recall} & {\textbf{Accuracy}} & {\textbf{F-Score}} \\ \hline
        0.5                & 0.532              & 0.899           & 0.867             & 0.668            \\
        0.6                & 0.710              & 0.872           & 0.928             & 0.783            \\
        0.7                & 0.814              & 0.852           & 0.949             & 0.833            \\
        0.8                & 0.829              & 0.812           & 0.947             & 0.820            \\
        0.9                & 0.853              & 0.691           & 0.937             & 0.766            \\ \hline
      \end{tabular}
      
      \label{tab:con_res}
      \vspace{+0.15cm}
    \end{table}
    
    \footnotetext{Positive: Unconnected, Negative: Connected}
    
    We also introduce a threshold $\lambda$, as we did in the ISAV measurements. If there is $\overline{rcv_2} / \overline{rcv_1} \ge \lambda$, we will infer that this IP address is unconnected with the vantage point B. 
    Figure \ref{fig:roc} is the receiver operating characteristic (ROC) curve of our method when we apply different $\lambda$ for $k$-time measurements. When $k \ge 4$, the area under the curve (AUC) stabilizes at 0.92, which demonstrates the capability of our method and suggests that four measurements have been sufficient to make the results stable and convincing. Table \ref{tab:con_res} lists the experimental results of our measurements with different $\lambda$ and $k=6$, revealing a relatively stable accuracy of over 90\%, precision and recall of over 80\%, which we believe is already very good for a theoretically impossible measurement task.

    \subsection{Limitations \& Challenges}
    
    \textbf{Estimation of RTTs.} The main challenge is how we control the interval between step 3 and step 6 so that the reflected packets sent by B (i.e., the ICMP Echo Replies) reach A at the same time as the probe packets sent in step 6 arrive. 
    In practice, we estimate the RTT between A and B first (very roughly!) based on their geolocation information \cite{jsacanycast} (using \verb|GeoLite2| \cite{geolite2}) and the triangle principle \cite{popovski2018wireless, phd1994}, where $d$ represents the distance between A and B, and $c$ is the speed of light:

    \begin{equation*}
      \left\{
                   \begin{array}{c}
                    \dfrac{d}{2c/3} \le \widehat{RTT_{AB}} \le \dfrac{d}{c/3} \\
                    \vert RTT_A - RTT_B \vert < \widehat{RTT_{AB}} < RTT_A + RTT_B  
                   \end{array}
      \right.
      \end{equation*}

    We randomly choose different value within this range as the estimated RTT for each measurement. We then set the time interval ($\Delta t$) between step 3 and step 6 to $\Delta t = (RTT_B - RTT_A + \widehat{RTT_{AB}})/2$ so that the reflected packets and our probe packets arrive simultaneously.
    Note that $\Delta t < 0$ is possible because we may do step 6 first instead of step 3.
    
    The RTT estimation does not need to be very accurate for two main reasons. First, in practice, a slightly longer $\Delta t$ is still fine because it is acceptable to let the reflected packets arrive a little earlier than the probe packets since the ICMP rate limiting lasts for a short period of time. Second, we randomly try different estimates in each measurement, so it would be sufficient if \textit{some} estimates were relatively accurate, which would result in a significant decrease in $\overline{rcv_2}$.
    However, it is unacceptable if our probe packets arrive earlier.
    Further analysis shows that there is usually an unexpectedly small RTT between the IP address that is misclassified in the evaluation and the vantage point B.
    An incorrect estimate of RTT is likely to trigger a premature or late ICMP rate limiting, posing an obstacle to our measurements. However, this challenge also reveals a novel approach for estimating the latency between two arbitrary nodes, which we leave for future work (\S \ref{sec:discussion2}).
    
    \textbf{Coverage of \Ys.} Our \Ys cover $\sim$30\% of BGP prefixes,  $\sim$50\% of ASes, and almost all countries. Usually, our main concern is the reachability between nodes in different prefixes, ASes, or countries. Therefore, we have theoretically $\sim$51\%, $\sim$75\%, or almost 100\% probability of finding at least one appropriate proxy \Y in the same prefix, AS or country as either of two targets, respectively. Considering that networks where we can't discover any \Ys may have very few active IPv6 hosts, the probabilities can be higher in our practical measurements. Actually, in terms of the numbers of ASes, the coverage of our \Ys is already better than all the existing censorship monitoring platforms \cite{iclab,globalDNSmanipulation,satellite, quack, censorglobalscale, ooni}.
    
    Note that generally, this method will not be affected by the ISAV deployed in the target networks because the spoofed source addresses of the packets we send in step 3 usually do not belong to the target networks, and thus will not be filtered by ISAV.

    \section{Discussion: ICMP Rate Limiting}
    \label{sec:discussion}
    This section provides a detailed discussion of ICMP rate limiting with respect to Internet-wide implementation, security and privacy risks, and possible mitigation measures, respectively.
    
    \subsection{How do Internet Nodes Implement ICMP Rate Limiting?}
    
    We perform a large-scale measurement of the implementation of ICMP rate limiting. We select 25,741 \Ys among 25,741 different longest-match BGP prefixes belonging to 8,834 ASs, focusing on \textit{breadth} rather than quantity, to provide a comprehensive view of ICMP rate limiting implementation.

    We aim at measuring the rate limiting of three types of ICMP messages: \textit{Time Exceeded} (by adjusting the hop limits), \textit{Destination Unreachable} (by sending ICMP Echo Requests to unreachable addresses, which are collected previously), and \textit{Echo Reply} (by sending ICMP Echo Requests to the \Ys directly).
    We also test whether the ICMP rate limiting is global by sending additional noise packets. Global, in this case, means limiting the rate of generation of ICMP error messages sent to all IP addresses, even if triggered by only one IP address. As we did in our ISAV measurements, we continuously measure the values of $rcv_1$ and $rcv_2$ for these three types of ICMP error (or informational) messages. As in the ISAV measurements, we let $N=50, M=100$ for ICMP error messages, and $N=M=500$ for ICMP Echo Replies.

    \begin{figure*}[htb]
      \centering
      \includegraphics[width=0.98\textwidth]{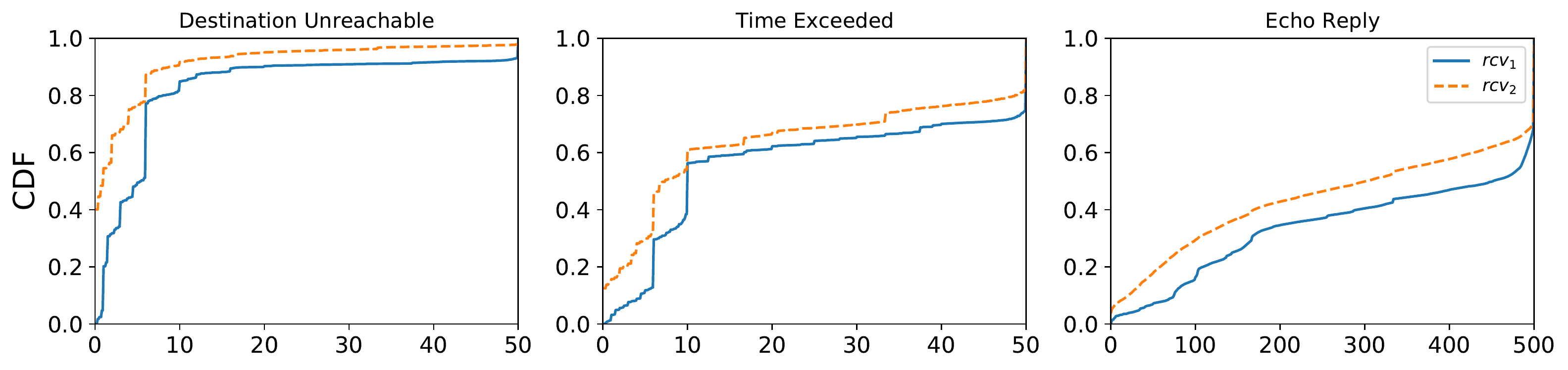}
      \caption{Distributions of $rcv_1$ and $rcv_2$ of Different Types of ICMP Messages}
      \label{fig:rlcdf}
    \end{figure*}

    Based on the measurement results (Figure \ref{fig:rlcdf}), we further calculate the percentage of global, strict, and loose ICMP rate limiting implementations (Table \ref{tab:rl_type}):
    \begin{itemize}
      \item \textbf{Global}: $\overline{rcv_2} < \lambda \times \overline{rcv_1}$ (we set $\lambda=0.6$ in practice). This kind of ICMP rate limiting can be well exploited by \X.
      \item \textbf{Strict}: $0.95\le\overline{rcv_1}\le1.05$. This kind is more secure because it is difficult (but still possible) to observe the difference before and after the rate limiting is triggered.
      \item \textbf{Loose}: $\overline{rcv_2} \ge 0.95N$, i.e., very loose (or even no) ICMP rate limiting is implemented, which can not be exploited as side channels but is vulnerable to ICMP flooding attacks.
    \end{itemize}

    \begin{table}[htb]
      \centering
      \caption{Percentages of Global, Strict, and Loose ICMP Rate Limiting Implementations}
      \begin{tabular}{cccc}
        \hline
        \textbf{Type}                    & \textbf{Global} & \textbf{Strict} & \textbf{Loose} \\ \hline
        {ICMP Destination Unreachable} & 72.16\%         & 15.46\%         & 2.41\%         \\
        {ICMP Time Exceeded}           & 38.84\%         & 1.94\%          & 21.03\%        \\
        {ICMP Echo Reply}              & 40.11\%         & 0.88\%          & 35.63\%        \\
        \hline
      \end{tabular}
      
      \label{tab:rl_type}
      \vspace{+0.2cm}
    \end{table}
    
    \textit{Findings.} We make the following findings:
    
    \begin{itemize}
      \item ICMP rate limiting is prevalent, with at least 65\%-98\% of the tested targets implementing significant ICMP rate limiting\protect\footnotemark. Of these, the rate limiting of ICMP Destination Unreachable is more stringent and has been observable in more than 97\% of cases with $N=50$. In contrast, the rate limiting of other two types is looser and more difficult to observe. Therefore, \X can make good use of ICMP Destination Unreachable without sending a large number of packets.
    
      \item Global ICMP rate limiting is indeed common, especially in the case of ICMP Destination Unreachable. We estimate that more than 50\% of Internet nodes enforce global rate limiting of all ICMP messages.
    
      \item The majority of tested Internet nodes ($>70\%$) implement global rate limiting of ICMP Destination Unreachable. Moreover, the rate limiting for ICMP Destination Unreachable is easy to trigger and observe, with only $\sim$18\% implementing strict or loose rate limiting, indicating that \X can be widely used for different \Ys distributed across the Internet.
    
    \end{itemize}
    
    \footnotetext{As we limit the amount of packets we send because of ethical concerns, some loose implementations may no longer be loose if $M$ and $N$ increase.}

    \subsection{Potential Risks}
    Internet standards keep striving to remove global things to protect from potential side channel-based attacks and measurements  \cite{rfc6274, rfc7739}. Researchers exploit global IPID counters to perform alias resolution \cite{midar, speedtrap}, stealthy scans \cite{idlescan}, and TCP hijacking \cite{tcphijacking}. In case that IPID counters are not global, global SYN caches are used as substitutes \cite{idleportscan, originalsyn}.
    Global ICMP rate limiting, though less harmful, can also be dangerous.
    By taking advantage of the large IPv6 address space, it is easy to induce IPv6 nodes to originate ICMP Destination Unreachable messages. Our measurements also confirm that the rate limiting of ICMP Destination Unreachable messages is easily observable and mostly global and thus can be well exploited as a side channel.
    Therefore, global ICMP rate limiting can be a good substitute for the well-known global IPID counter for side channel-based measurements. After all, the IPID field has been removed from the IPv6 fixed header. The only difference is that we observe the state of ICMP rate limiting, not the increment of the IPID. It is often easier to observe the state of ICMP rate limiting in the presence of noise.
    In addition to our work, the side channels of ICMP rate limiting also reveal router configurations \cite{asByRL} and open ports \cite{ccs2020}. All these works demonstrate that ICMP rate-limiting side channels may lead to security and privacy issues.

    \subsection{Mitigation Measures}
    We further discuss some possible mitigation measures to prevent ICMP rate limiting from being exploited as side channels.
    
    \textbf{Strict or Non-global ICMP Rate Limiting is Recommended.}
    Non-global rate limiting is an intuitive and effective solution. For example, rate limiting for ICMP port 444 unreachable will not interfere with the rate limiting state that generates ICMP port 445 unreachable, and rate limiting for ICMP messages sent to source A will not interfere with the rate of ICMP messages sent to source B. However, non-global rate limiting is difficult to implement and deploy because too many rate limiting counters need to be maintained (e.g., token buckets as recommended by the RFC \cite{rfc4443}). An easy-to-deploy alternative solution that we recommend is to implement strict ICMP rate limiting, i.e., to send only one ICMP message regardless of how many packets are received in a short period of time. \X and other efforts to exploit ICMP rate limiting \cite{asByRL, ccs2020} rely on receiving a different number of ICMP messages before and after triggering ICMP rate limiting. Even though still global, strict rate limiting makes the differences much less observable. However, strict ICMP rate limiting cannot cope with bursty traffic and is therefore not recommended by the RFC \cite{rfc4443}. Thus, there may be a trade-off, and it is still difficult to find a perfect solution. Side channels of ICMP rate limiting may be exploitable for a long time to come.

    \textbf{ICMP Destination Unreachable Should be Restricted.} It is easy to find an unreachable IP address in such a large IPv6 address space, so it is also easy to induce IPv6 nodes to initiate ICMP Destination Unreachable messages. The node initiating an ICMP Destination Unreachable message exposes itself, which can then be exploited. Just like the process of \Y discovery, actually we do not have any active IPv6 addresses at first. However, by sending ICMP Echo Requests to those fabricated destination IP addresses, we can discover a great many active IPv6 addresses by receiving ICMP error messages. Then, they can even be used as our ``vantage points''! Therefore, allowing IPv6 nodes to generate ICMP Destination Unreachable messages without any restrictions will be dangerous. 
    For example, when a router receives a series of packets destined for very strange destination addresses within its subnet (especially if these packets are sent from a remote network), it may be a safer choice to ignore them than to initiate ICMP destination unreachable messages for each packet.
    Especially, we also recommend not initiating ICMP error messages in response to ICMP Echo Replies\protect\footnotemark.
    
    \footnotetext{Though IPv6 nodes are not allowed to originate ICMP error messages as a result of receiving ICMP error messages according to the RFC \cite{rfc4443}, ICMP Echo Replies are not ICMP error messages.}

    \section{Limitations and Future Work}
    \label{sec:discussion2}

    \X makes use of remote routers as ``vantage points'' to perform active measurements via ICMP rate limiting side channels. Admittedly, there are still some limitations of \X and our work. We look forward to improving our work in the future from following aspects:

    \subsection{More \X-based Measurement Applications}

    We present two typical applications of \X in this paper, which seems to be kind of limited, but we believe that our work is just the beginning. There are still many other measurements or attacks that can be performed based on \X to be discovered. Note that it is also possible to send other packets (e.g., TCP-SYN, UDP, DNS queries, NTP, etc.) instead of \verb|ping| as probe packets. \textbf{All packets that the destination needs to reply to can be used as probe packets, and we can measure the reachability of these packets based on \protect\X}.

    \begin{figure}[htbp]
      \centering
      \includegraphics[width=0.48\textwidth]{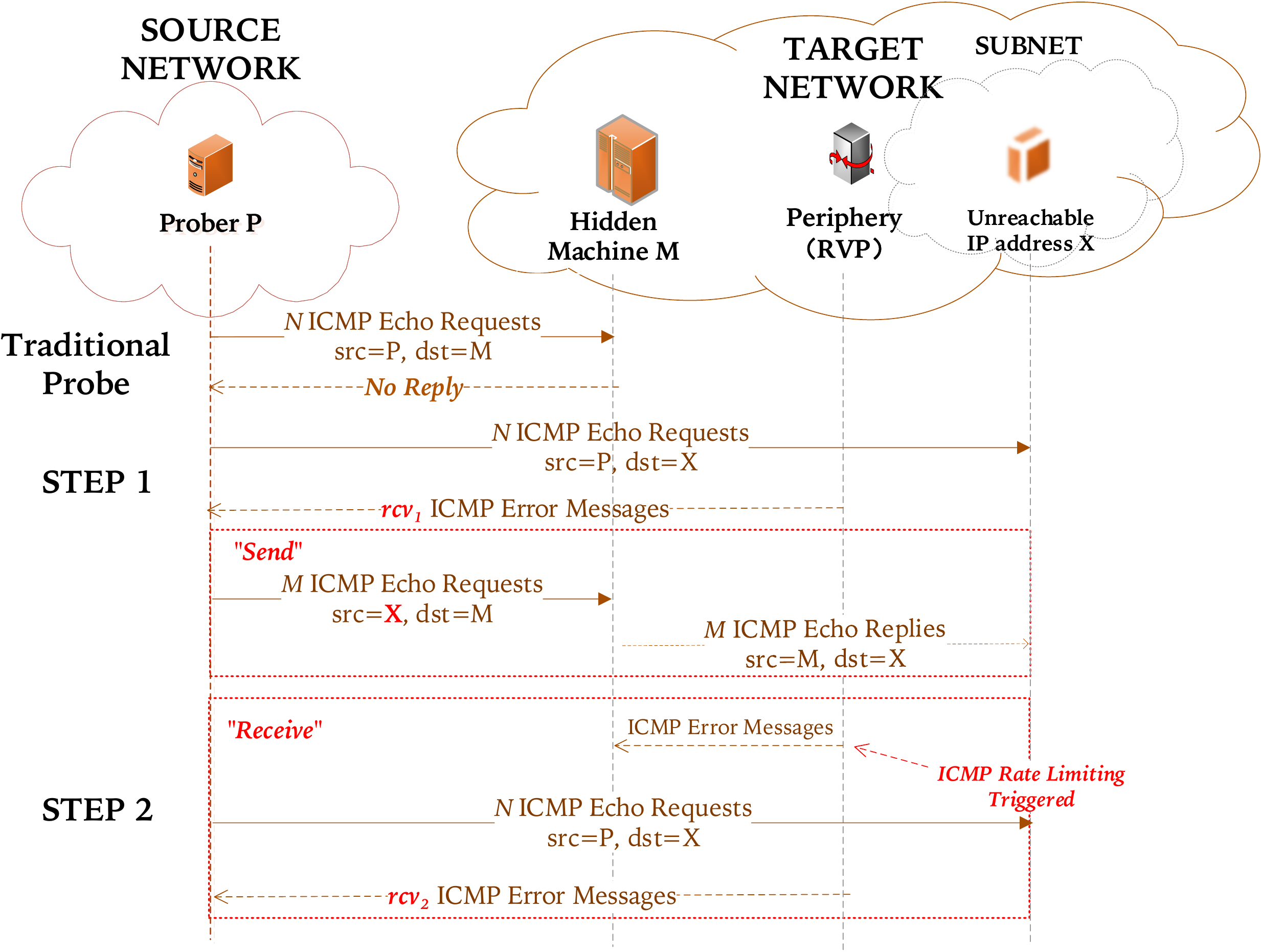}
      \caption{Novel Idea based on \X to Discover Hidden Machines. The existence of the hidden machine can be identified by comparing $rcv_1$ and $rcv_2$.}
      \label{fig:hidden}
    \end{figure}

    For instance, Figure \ref{fig:hidden} shows a novel idea we have just come up with based on \X to discover hidden machines in remote network. Just like the famous TCP idle scan \cite{idlescan} and its variations\cite{idleportscan,originalsyn, tcphijacking}, this new idea exploits a zombie machine (i.e., \Y) to discover hidden machines that only respond to specific devices in its own networks (e.g., only reply to the \verb|ping| sent from the network it belongs to). This is common, especially for some local servers, such as local DNS resolvers. We will deeply dive into this new idea in the future, and there may also be other novel and also interesting ideas based on \X like that to be found.

    \subsection{Demystifying ICMP Rate Limiting Behaviors} 
    This paper presents an Internet-wide measurement on ICMP rate limiting implementations, but we think there is more to take a deep dive into. For example, we just identify ICMP rate limiting in a very intuitive (but we believe effective) way: by comparing the number of replies we received. We are going to focus on the characterization of behaviors and the implementations of ICMP rate limiting, which may help us identify rate limiting more accurately and efficiently with fewer packets to be sent. 

    \subsection{Accurate Latency Estimation between Remote IPv6 Nodes} 
    As mentioned earlier in \S \ref{sec:measuring reachability}, by checking the state of ICMP rate limiting, we can know whether the packet from step 4 or step 6 arrives first at A. Thus, by controlling $\Delta t$ between step 3 and step 6 and checking the state of ICMP rate limiting on A, we can estimate the RTT between A and B. This can also be considered to be another measurement application of \X.

    \subsection{IPv4 Applications}
    \X faces challenges in IPv4 networks for two reasons: 1) ICMP rate limiting is not required \cite{rfc792} (though many may implement\cite{asByRL, characterizeICMP, detectRL}, not that pervasive\cite{asByRL, beholder}) in IPv4 networks; 2) It's also more difficult to find unreachable addresses to induce possible \Ys to initiate ICMP Destination Unreachable, which is the type of ICMP error message that we mainly exploit in \X. How to make \X well applicable to IPv4 networks is what we may consider in the future.

    \section{Conclusion} 
    \label{sec:conclusion}
    In this paper, we focus on ICMP rate limiting side channels. We propose a novel technique, \X, and apply it to two difficult and theoretically impossible measurement tasks: measuring ISAV deployment and reachability between remote Internet nodes from only one local vantage point. In addition, we measure the implementation of ICMP rate limiting, reveal the security and privacy risks of existing ICMP rate limiting implementations, and provide possible mitigation measures. We will further study on ICMP rate limiting and its side channels in the future.

    \section{Ethical Considerations}
    Before performing our measurements, we looked through some key guidelines for Internet measurements\cite{ethics16, ethics17, menloreport}.
    Since our department does not have an Institutional Review Board, we consulted the academic board of our department. Feedback from the academic board mainly concerned the possible effect on the target devices when triggering ICMP rate limiting.
    We replied that: 1) ICMP rate limiting is a required and basic function of IPv6 nodes, and triggering ICMP rate limiting is also common on the real Internet, generally not considered harmful; 2) our measurements will never trigger continuous ICMP rate limiting on the same target (see below); 3) there already existed several published papers exploiting the ICMP rate limiting mechanism\cite{ccs2020, asByRL}, particularly, lab experiments in \cite{asByRL} show that ICMP rate limiting does not have a discernible negative effect on the devices. The academic board approved our study after serious consideration. In practice, we have taken into account following aspects for ethical considerations in our measurements. 
    
    \subsection{Anonymity}

As many previous work on ISAV, we totally ensure the anonymity of prefixes and ASes we measured to prevent those vulnerable-to-spoofing networks from being attacked by spoofing-based cyberattacks. We also do not make the IP addresses of \Ys we discovered public to the community.

\subsection{Relatively Harmless Probes}
    We only send ICMP Echo Requests (i.e., \verb|ping|) in our measurements. Compared with other types of scans like port scans and sending DNS queries, sending ICMP Echo Requests is relatively harmless. We prevent probing one network in succession in all parts of our measurements. For instance, in the process of \Y discovery, we randomized the probing sequence by the multiplicative group of integers modulo $n$ \cite{gauss,ZMap} like many existing high-speed probers \cite{ZMap,beholder, flashroute}.
    Network administrators can easily contact us by the e-mail address in the \verb|WHOIS| database or the reverse DNS record. We received 5 complaints during our measurements, and their networks (the whole ASes) are excluded from our subsequent measurements.

    \subsection{Limiting the Amount of Packets}
    It is inevitable to send many packets to trigger ICMP rate limiting.
    However, we strictly limit the amount of packets we send.
    We only send 50 ICMP Echo Requests to check the state of
    ICMP rate limiting (500 for the rate limiting of ICMP Echo Replies).
    If the ICMP rate limiting is relatively loose (e.g., receive 50/50 ICMP error messages),
    we will not
    further increase the amount of our packets, even though we know this will help trigger more
    obvious ICMP rate limiting.
    Instead, we try another \Y.
    Similarly, our noise packets will be no more than 100 (500 for ICMP Echo Replies),
    even though more noise packets may help us observe more obvious difference.
    While other studies exploiting ICMP rate limiting like \cite{asByRL} usually send thousands of packets per second, our measurements send much fewer packets.

    \subsection{Preventing Continuous ICMP Rate Limiting}
    Our measurements will \textbf{never} trigger ICMP rate limiting on one Internet node
    in succession. In \Y discovery, probes in random order prevent
    ICMP error messages from being continuously sent from one periphery;
    in our ISAV measurements, as introduced before,
    we will not measure the values of $rcv_1$, $rcv_2$ and $rcv_3$ of one network without break,
    instead, we measure $rcv_1$ of the first network, and then measure $rcv_1$
    of the second network; in our measurements of reachability,
    we use different \Ys in rotation, and with a relatively long interval of five minutes;
    when measuring ICMP rate limiting implementations, we also prevent continuous ICMP rate limiting
    as we did in the ISAV measurements.
    In our measurements, due to our randomness, after rate limiting is triggered on an RVP (which we find usually lasts for a very short time), there is a long time before rate limiting is triggered again. Thus, the duration of rate limiting is a really tiny fraction compared to their normal operation time and does not interfere with their normal function.

    \subsection{Previous Laboratory Experiments} 
    We refer to previous work on rate limiting, e.g., \cite{asByRL}, where they conducted real laboratory experiments on routers that sent more packets and triggered more severe rate limiting than our work. They found that rate limiting does not lead to destructive problems or high CPU usage, so our approach of much fewer probe packets and larger intervals does not severely affect networks. As a basic and required function of IPv6 nodes, we believe ICMP rate limiting will not lead to a disruptive impact on either the data plane or the control plane of the target device.

    \subsection{Real Internet Experiments} 
    We request access to several edge routers in our campus network from the network administrators and monitor the changes in CPU usages, memory usages when we use them as \Ys in our ISAV and reachability measurements. However, we cannot observe \textit{any} observable changes in CPU usages or memory usages on them, even after we increase the number of packets we send by a factor of 2 or 3. We also cannot observe any abnormal behaviors and we believe our measurements do not interfere with routers' basic functions. We think the reasons may include but is not limited to: 1) For every router, our measurement last only a short time over a long period of time (usually much less than 1 second), and routers usually cannot provide real-time performance monitoring with a granularity of less than 1 second to capture the changes. 
    2) The number of packets we send is also very small compared with thousands or even millions of packets the router forwards per second. 
    3) Common token-bucket implementations of ICMP rate limiting, just as recommended by the RFC \cite{rfc4443}, are really resource-saving and do not result in obvious increase of CPU and memory usages.

    \section*{Acknowledgements}
    The authors would like to thank Erik Rye et al. \cite{periphery, prefixrotation} and Xiang Li et al. \cite{xiangli} for their previous contributions to the discovery of IPv6 peripheries, which are crucial preliminary to our work. Previous research on ICMP rate limiting side channels by Man et al. \cite{ccs2020} and Vermeulen et al. \cite{asByRL} are also great inspirations for us. This work is supported by the National Key Research and Development Program of China under Grant No. 2018YFB1800200 and Beijing Natural Science Foundation under Grant No.4222026. Lin He is the corresponding author of this paper: \href{mailto:he-lin@tsinghua.edu.cn}{he-lin@tsinghua.edu.cn}.

\bibliographystyle{IEEEtranS}
\bibliography{IEEEabrv, sample-base}

\end{document}